\documentclass[12pt]{iopart}

\usepackage{graphicx}
\usepackage{dcolumn}
\usepackage{bm}
\usepackage{amsmath}
\usepackage{array,multirow,graphicx}
\usepackage{float}
\usepackage[caption=false,subrefformat=parens,labelformat=parens]{subfig}
\usepackage{lipsum}
\usepackage{float}
\usepackage{mathtools} 
\usepackage[utf8]{inputenc}
\usepackage[T1]{fontenc}
\usepackage{url}
\usepackage[table]{xcolor}
\usepackage{verbatim}
\usepackage{eufrak}
\usepackage{array}
\usepackage[colorlinks,linkcolor=blue,anchorcolor=blue,citecolor=blue,urlcolor=blue]{hyperref}
\usepackage{url}
\hypersetup{breaklinks=true}
\usepackage{booktabs}
\DeclareMathAlphabet\mathbfcal{OMS}{cmsy}{b}{n}
\begin{document}

\title{New Angular Momentum Conservation Laws for Gauge Fields in QED}

\author{Farhad Khosravi$^{1,3}$, Li-Ping Yang$^{2,3}$, Pronoy Das$^3$ and Zubin Jacob$^{3,*}$}
\address{$^1$1QBit, Evergreen Building, Vancouver, British Columbia V6E 4B1, Canada}
\address{$^2$Center for Quantum Sciences and School of Physics, Northeast Normal University, Changchun 130024, China}
\address{$^3$Elmore Family School of Electrical and Computer Engineering, Birck Nanotechnology Center, Purdue University, West Lafayette, IN 47907, USA}%

\ead{zjacob@purdue.edu*}

\vspace{10pt}
\begin{indented}
\item[]March 2024
\end{indented}

\begin{abstract}
Quantum electrodynamics (QED) deals with the relativistic interaction of bosonic gauge fields and fermionic charged particles. In QED, global conservation laws of angular momentum for light-matter interactions are well-known. However, local conservation laws, i.e. the conservation law of angular momentum at every point in space, remain unexplored. Here, we use the QED Lagrangian and Noether's theorem to derive a new local conservation law of angular momentum for Dirac-Maxwell fields in the form of the continuity relation for linear momentum. We separate this local conservation law into four coupled motion equations for spin and orbital angular momentum (OAM) densities. We introduce a helicity current tensor, OAM current tensor, and spin-orbit torque in the motion equations to shed light on on the local dynamics of spin-OAM interaction and angular momentum exchange between Maxwell-Dirac fields. We elucidate how our results translate to classical electrodynamics using the example of plane wave interference as well as a dual-mode optical fiber. Our results shine light on phenomena related to the spin of gauge bosons.

\end{abstract}

%
%
%
%
%

\section{Introduction}

\begin{figure*}[t!]
     \centering
        \subfloat[\label{Fig:Schematic_Old}]{
        \includegraphics[width=0.38\linewidth]{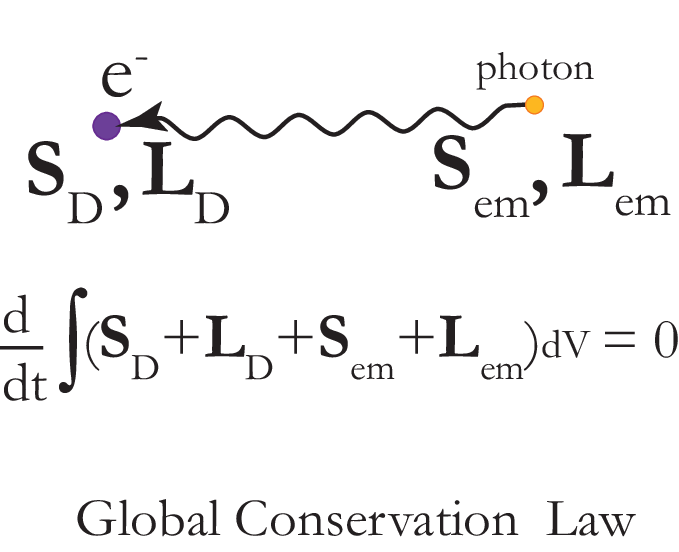}} \hspace{4mm}
        \subfloat[\label{Fig:Schematic_New}]{
        \includegraphics[width=0.52\textwidth]{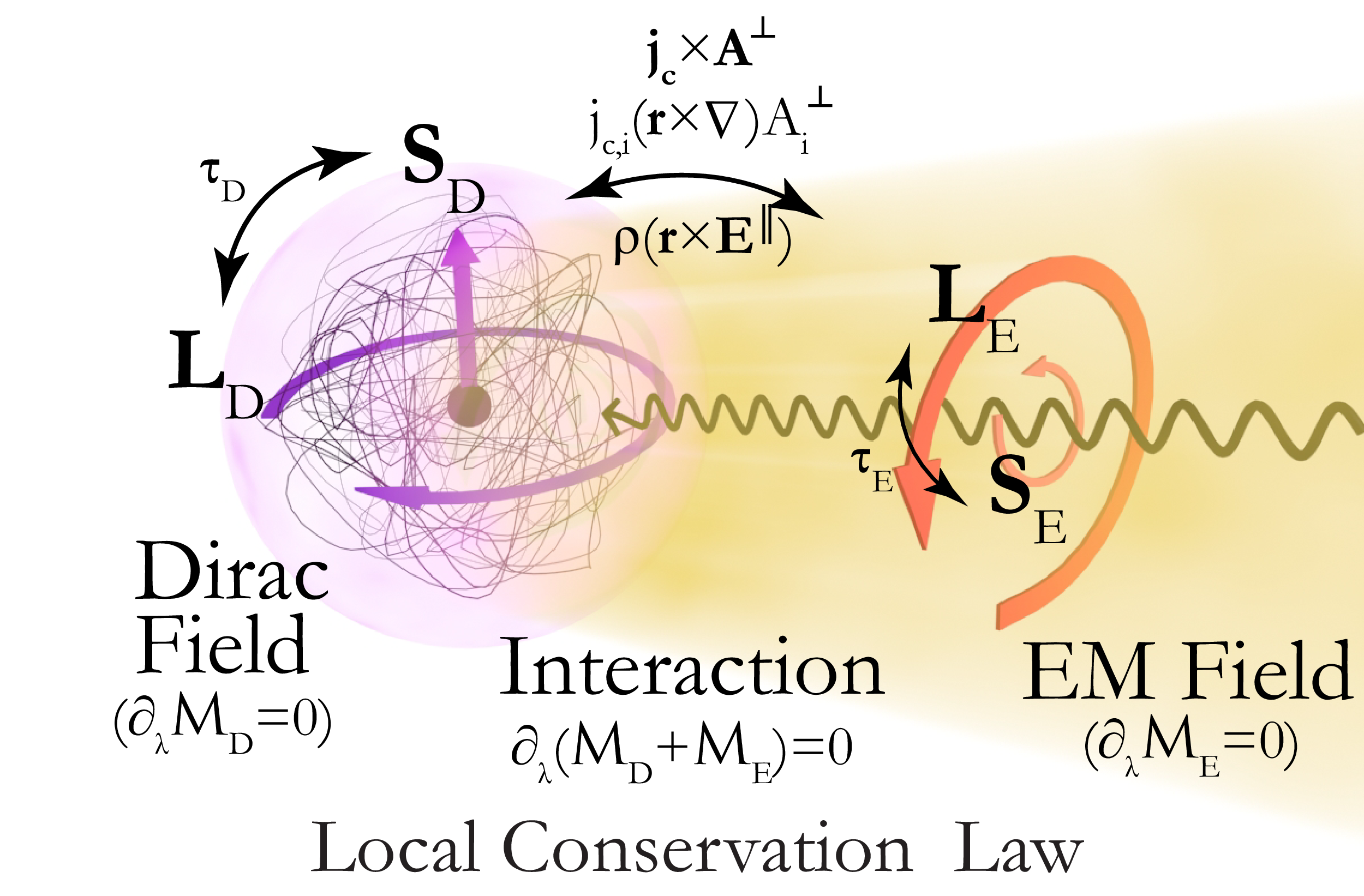}} \par \vspace{4mm}
        \subfloat[\label{Tab:Quantities}]{
        \resizebox{\textwidth}{!}{%
        \begin{tabular}{|c|c|c|c|c|c|c|}
        \hline
        Behavior & Scalar & \multicolumn{3}{c|}{Vector} & \multicolumn{2}{c|}{Tensor} \\ \hline
        \multirow{2}{*}{Physical quantity} & Chirality / & Spin Angular & Orbital & Spin-orbit &  Helicity & Orbital Angular \\
        & Helicity & Momentum & Angular Momentum & Torque & Current & Momentum Current \\ \hline
        \textbf{Dirac Field} & $\frac{\hbar}{2}\psi^\dagger \gamma^5 \psi$ & $\frac{\hbar}{2}\psi^\dagger \bm \Sigma \psi$ & $\mathcal{R}\{\psi^\dagger (\bm r \times \bm p_\parallel)\psi\}$ & \cellcolor{yellow!50} $- c \mathcal{R}\left\{ \bar{\psi}\left(\bm \gamma \times\bm p_\parallel\right)\psi \right\}$ & $0$ & \cellcolor{yellow!50} $\mathcal{R}\left\{ \bar{\psi}\gamma_i (\bm r\times \bm p_\parallel)_j \psi \right\}$ \\ \hline
        \multirow{2}{*}{\textbf{Maxwell Field}} & & \multirow{2}{*}{$ \epsilon_0 (\bm E^\bot \times \bm A^\bot)$} & \multirow{2}{*}{$\epsilon_0 \left[ E_i^\bot (\bm r \times \nabla)A_i^\bot \right]$}  & \cellcolor{yellow!50}$\frac{1}{\mu_0} (\bm B \cdot \nabla)\bm A^\bot$ &   \cellcolor{yellow!50} & \cellcolor{yellow!50}$-\frac{1}{\mu_0} \left[\varepsilon_{ikl} B_k (\bm r \times \bm\nabla)_j A_l^\bot \right] $ \\
        & \multirow{-2}{*}{$\frac{1}{\mu_0 c}\left(\bm A^\bot \cdot \bm B\right)$} & & & \cellcolor{yellow!50} $-\epsilon_0 \left(\frac{\partial \bm E^\parallel}{\partial t} \times \bm A^\bot\right)$  & \cellcolor{yellow!50}\multirow{-2}{*}{ $-\frac{1}{\mu_0}( A^\bot_i B_j)$} & \cellcolor{yellow!50} $ - \epsilon_0 A_i^\bot \left(\bm r \times \frac{\partial \bm E^\parallel}{\partial t}\right)_j$ \\ \hline
        \end{tabular}}
        }
    \caption{Conservation laws of angular momentum in light-matter interacting systems (a) conventional conservation of global total photonic and electronic angular momentum. This conservation law applies to closed systems and does not include angular momentum exchange due to near-field interactions. (b) Local conservation of angular momentum applicable to all regions of interaction. The conventional conservation of sum of angular momenta is replaced by the conservation Eq.~(\ref{Eq:Conservation_Law_02}). (c) Table of quantities defined in this paper, representing the spatial and temporal densities of the scalar, vector, or tensor observables pertinent to angular momentum. Highlighted quantities are the new terms defined in this paper. The local conservation of angular momentum equation (Eq.~(\ref{Eq:Conservation_Law_02})) connects the well-known terms such as spin density, OAM density, helicity and chirality to these newly defined quantities.}
    \label{Fig:Schematic}
\end{figure*}

Global conservation laws for angular momentum are well known in quantum electrodynamics. However, the spatial distribution of angular momentum (AM) i.e. AM density is relatively unexplored in QED. Here, the gauge fields take center stage but no local conservation laws are known for the U(1) gauge field in QED. Fig.~\ref{Fig:Schematic_Old} highlights the difference between global and local conservation of total angular momentum. It also highlights unique physical quantities related to angular momentum for both Dirac fields and Maxwell fields. We emphasize that  in the presence of sources,  a comprehensive, gauge-independent, locally applicable conservation equation becomes essential in understanding the local dynamics of the angular momentum (Fig.~\ref{Fig:Schematic_New}).

On the other hand, the developments in classical electrodynamics have focused on both global and local angular momentum. The local approach towards field quantities such as helicity, chirality, and angular momentum has proven useful in the study of conservation laws as well as geometrical properties of EM fields  \cite{Zhou2021,Carlos2021,crimin2019optical,bliokh2013dual,berry2009optical,nienhuis2016conservation,fernandez2013electromagnetic,tang2010optical,philbin2013lipkin}. However, the procedure employed in these derivations is based on the duality symmetry of the source-free EM Lagrangian. Since the duality symmetry is only maintained in source-free regions, this form of Lagrangian neglects any form of interaction with fermionic fields and cannot be employed to present a local conservation law for the angular momentum of gauge fields in QED. These properties are captured in the manifestly covariant construction of the Dirac equation \cite{greiner1990relativistic}. Connections between the fermionic field of the Dirac equation and the bosonic fields of Maxwell's equations is the focus of Dirac-Maxwell correspondence where the relativistic parallels between photons and electrons are studied \cite{barnett2014optical,bialynicki1996v}. These studies show that, although different in nature, electronic and electromagnetic fields can exhibit many analogous properties. This correspondence is evident in phenomena such as spin-momentum locking which emerges in the Dirac equation \cite{khosravi2019dirac} as well as the Maxwell's equations \cite{van2019nonlocal,van2019unidirectional}.

In this paper, we study angular momentum of light using the Lorentz symmetry of the QED Lagrangian and local $U(1)$ gauge invariance \cite{greiner1996gauge}. With the application of Noether's theorem \cite{noether1971invariant}, we find unique local conservation laws pertinent to the angular momentum of the gauge field interacting with fermionic fields. We show that, in the near-field, the electronic angular momentum can be transferred not only to the optical angular momentum, but also to other field quantities that represent angular momentum current \cite{barnett2001optical}. These extra terms include electromagnetic helicity density \cite{calkin1965invariance}, fermionic chirality density \cite{greiner1990relativistic}, electromagnetic \textit{helicity current tensor}, as well as the electromagnetic and electronic \textit{orbital angular momentum (OAM) current tensors}. We further study these conservation laws and angular-momentum-carrying terms for the classical electromagnetic solutions of a dual-mode optical fiber. This connection between QED and classical electrodynamics will be of interest for future experiments related to OAM and SAM of light.

\section{Noether's theorem, QED Lagrangian, and Lorentz transformation}

An important problem in QED relates to the angular momentum of the gauge field. Particularly, the goal is to decompose the total angular momentum of gauge bosons into spin and orbital contributions ~\cite{leader2014angular}. In early work, in both Belinfante's~\cite{belinfante1940current} and Ji's~\cite{ji1997gauge} decompositions, only the total angular momentum of photons has been explored in detail. Jaffe and Manohar split the photonic angular momentum into spin and orbital parts~\cite{jaffe1990g1}. However, their decomposition does not satisfy the gauge-invariant requirement. By splitting the vector potential into transverse and longitudinal parts, Chen et al.~\cite{chen2008spin} and Wakamatsu~\cite{Wakamatsu2010gauge} propose gauge-invariant decompositions of QED angular momentum. However, these existing decompositions have not resolved the conflict between the gauge-invariant requirement and canonical angular momentum commutation relations. Recently, this problem has been solved by quantizing the U(1) gauge field in the Lorenz gauge and merging the spin and OAM of virtural photons with the OAM of Dirac fermions~\cite{yang2020quantum,Das2024arxiv}.

The application of Noether's theorem to the Dirac Lagrangian has been used in the Quantum Chromodynamics (QCD) and QED communities to derive the expressions for the angular momentum of the nuclei and the gauge fields \cite{ji1997gauge,chen2008spin}. In QCD and QED interactions, conservation of the total integrated angular momentum suffices to describe the angular momentum transfer in scattering processes. On the other hand, for a fast moving electron, its spin and momentum degrees of freedom are highly correlated~\cite{Foldy1950on}. Explorations of a generalized definition for the spin operator of an electron as well as for gauge bosons is still a problem of widespread interest ~\cite{Terno2003two,Fujikawa2014spin} in QED and QCD \cite{leader2014angular,ji1997gauge,chen2008spin}. While the focus has mainly been on global conservation laws for angular momentum, our goal in this paper is to focus on local space dependent AM density effects. In nanophotonic and condensed matter systems, atoms can interact locally with an external EM field. Therefore, for such systems, a local conservation equation for the angular momentum density is necessary in the realization of naonscale applications. 


In this paper, we start with the real (also called symmetrized) Lagrangian density of a Dirac field coupled to EM field~\cite{greiner2013field},
\begin{equation}\label{Eq:Lagrangian}
    \mathfrak{L} = \Bar{\psi}\left[ c \gamma^\mu \left(\frac{1}{2}i\hbar  \overleftrightarrow{\partial}_\mu - e A_\mu \right) - mc^2
    \right]\psi - \frac{1}{4\mu_0}F_{\mu\nu}F^{\mu\nu},
\end{equation}
where $\bar{\psi} = \psi^\dagger \gamma^0$ and $\gamma^\mu$ are the gamma matrices, $F_{\mu\nu}$ the electromagnetic tensor, $A_\mu = (\phi/c,-\bm A )$ the electromagnetic four-potential, $\psi$ Dirac fields, $\hbar$ Planck's constant, and $\mu_0$ the vacuum permeability (Appendix). Note that summation over repeated indices is assumed throughout this paper. By applying Noether's theorem~\cite{noether1971invariant}, we obtain the conserved current for the given Lagrangian of Eq.~(\ref{Eq:Lagrangian}) by finding,
\begin{equation}\label{Eq:Noether_Currents}
    \mathcal{M}^\lambda = \frac{\partial \mathfrak{L}}{\partial (\partial_\lambda \psi)}\delta \psi + \delta \bar{\psi}\frac{\partial \mathfrak{L}}{\partial(\partial_\lambda \bar{\psi})} + \frac{\partial \mathfrak{L}}{\partial\left( \partial_\lambda A^\mu \right)} \delta A^\mu + \mathfrak{L}\delta x^\lambda.
\end{equation}
Lorentz transformations are defined as the coordinate transformations such that the coordinates transform as $x^\mu \to {\Lambda^\mu}_\nu x^\mu$, with 
\begin{equation}\label{Eq:Lorentz_Transformation}
    {\Lambda^\mu}_\nu = e^{-\frac{i}{2} \omega_{\kappa \sigma} {(\hat{S}^{\kappa \sigma})^\mu}_\nu}, \quad 
\end{equation}
where ${(\hat{S}^{\kappa \sigma})^\mu}_\nu = i\left( \eta^{\kappa\mu} {\eta^\sigma}_\nu - \eta^{\sigma\mu} {\eta^\kappa}_\nu \right) $ are the generators of rotation and boost in the four-dimensional real space \cite{kleinert2016particles} and $\omega_{\kappa\sigma}$ are the rotation and boost parameters. It can be shown that, under the Lorentz transformation of Eq.~(\ref{Eq:Lorentz_Transformation}), the Maxwell and Dirac fields change as,
\begin{equation}\label{Eq:Fields_Transformations}
    \delta A^\mu(x) = -\frac{i}{2} \omega_{\kappa \sigma} {( \hat{M}_\text{em}^{\kappa \sigma})^\mu}_\nu, \quad \delta\psi(x) = -\frac{i}{2} \omega_{\mu \nu}\hat{M}_\text{D}^{\mu\nu} \psi(x) 
\end{equation}
where $\hat{M}_\text{em}$ and $\hat{M}_\text{D}$ are the angular momentum operators of the Dirac and Maxwell's fields, respectively (Appendix). Specializing to the rotations ($\mu,\nu = i,j = 1,2,3$), we obtain the
general continuity equation associated with rotational symmetry of the QED Lagrangian \cite{kleinert2016particles}(Appendix),
\begin{equation}\label{Eq:Conservation_Law_01}
\begin{split}
    \partial_\lambda \mathcal{M}^{ij,\lambda} 
    = &     \partial_\lambda \left( \mathcal{S}^{ij,\lambda}_\text{D} + \mathcal{L}^{ij,\lambda}_\text{D} +  \mathcal{S}^{ij,\lambda}_\text{em} + \mathcal{L}^{ij,\lambda}_\text{em} \right) = 0,
\end{split}
\end{equation}
where $\mathcal{M}^{ij,\lambda}$ is the total angular current tensor. This tensor can be split into spin $\mathcal{S}^{ij,\lambda}_\text{D(em)}$ (related to the rotation of the internal degrees of freedom) and OAM $\mathcal{L}^{ij,\lambda}_\text{D(em)}$ (related to the coordinate dependence of the fields) parts of the Dirac and EM fields, respectively (Fig.~\ref{Fig:Schematic}). For non-interacting fields, we obtain separate continuity equations for the Dirac and EM fields. Note that the roman indices take the values $i,j = 1,2,3$ while the Greek indices are $\lambda = 0,1,2,3$. 

The time-components ($\lambda=0$) of these four angular momentum tensors give the common spin and OAM densites for the Dirac and EM fields, which are generally gauge dependent. The electronic part of the spin and OAM respectively are $\mathcal{S}^{ij,0}_\text{D}=\varepsilon_{ijk}\hbar\psi^\dagger \Sigma_k \psi/2$ ($\varepsilon_{ijk}$ is the Levi-Civita symbol) and $\mathcal{L}^{ij,0}_\text{D} = -i\varepsilon_{ijk}\hbar \mathcal{R}\left\{ \psi^\dagger \left( \bm r\times \nabla \right)_k\psi \right\}$, where $\bm \Sigma$ is the spin operator in Dirac equation and $\mathcal{R}\{\cdots\}$ takes the real part of its argument. The EM part of the spin and OAM, on the other hand, are $\mathcal{S}^{ij,0}_\text{em} =  \varepsilon_{ijk}(\epsilon_0 \bm E\times \bm A)_k$ and $\mathcal{L}^{ij,0}_\text{em} = \varepsilon_{ijk} \left[\epsilon_0 E_l (\bm r\times \nabla)_k A_l \right]$, respectively (Appendix), where $\epsilon_0$ is vacuum permittivity. An important observation is that OAM densities of the Dirac fields as well as the spin and OAM of the EM field are gauge dependent and thus do not represent observable physical quantities in a local frame. In fact, one can see that most of the terms in the angular momentum tensors $\mathcal{S}^{ij,\lambda}_{\rm D(E)}$ and $\mathcal{L}^{ij,\lambda}_{\rm D(E)}$ are gauge dependent. Due to this problem, we rearrange the expressions in these four tensors to write the conservation law in Eq.~(\ref{Eq:Conservation_Law_01}) in terms of gauge-independent and physically observable quantities. 

\section{Local conservation law of angular momentum}
The essence of Noether's theorem in field theory lies in the local conservation law with a general form $\partial_{\mu}j^{\mu}=0$. The global charge $Q\equiv\int d^3x j^0(x)$ conserves only if all fields vanish at spatial infinity, i.e., the vanishing boundary condition. As we mentioned above, the local conservation law of angular momentum has not received the attention it deserves both in experiments and in theory. We note that the density of the conserved charge in the original continuity equation (\ref{Eq:Conservation_Law_01}) of angular momentum is a rank-2 tensor and the corresponding current is a rank-3 tensor. This local conservation law is extremely difficult to apply to practical systems and almost impossible to verify in experiments. In this section, we re-express this relation as a continuity equation of a vector density similar to the local conservation law of linear momentum. The corresponding current reduces to a rank-2 tensor akin to the Maxwell stress tensor. We then go one step further by splitting our local conservation law into four coupled motion equations, which describe the spin-OAM interaction and angular momentum exchanging between Maxwell-Dirac fields.

By splitting the electromagnetic vector potential into transverse and longitudinal parts, $\bm A = \bm A^\bot + \bm A^\parallel$, defined as $\nabla\cdot\bm A^\bot = 0$ and $\nabla\times \bm A^\parallel = 0$~\cite{cohen1997photons}, X.-S. Chen \textit{et al} defined the gauge-independent angular momentum densities for both the Dirac field and EM field \cite{chen2008spin}. However, a gauge-independent form for the continuity equation has not been addressed. Here, following the same approach used in Ref.~\cite{chen2008spin}, we obtain a standard (vector) continuity equation for the angular momentum density of the combined system (see more details in the Appendix, \begin{equation}\label{Eq:Conservation_Law_02}
    \frac{\partial M_j}{\partial t} + \nabla_i {T}_{ij} = 0
\end{equation}
where $M_j$ are the components of the angular momentum density vector, written as
\begin{align}\label{Eq:TAM}  
\bm M  = & \frac{\hbar}{2}(\psi^\dagger \bm \Sigma \psi) + \mathcal{R}\left\{\psi^\dagger \left(\bm r \times \bm p_\parallel \right) \psi\right\} \\ \nonumber
& + \epsilon_0 (\bm E^\bot \times \bm A^\bot) +  \epsilon_0 \left[ E_i^\bot (\bm r \times \nabla) A_i^\bot \right],
\end{align}
with gauge-independent transverse vector potential $\bm A^\bot$ and gauge-independent electron momentum operator $\bm p_\parallel = \left( -i\hbar \nabla - e \bm A^\parallel\right)$. The first two terms in Eq.~(\ref{Eq:TAM}) are the spin and OAM densities of the electron, while the last two terms are the gauge-independent spin and OAM densities of the EM field~\cite{leader2014angular,chen2008spin}. This shows that only the transverse part of the vector potential contributes to the physically observable spin and OAM of EM field.

The angular momentum current tensor, $T_{ij}$, is a second rank tensor, similar to the Maxwell stress tensor (EM momentum current)~\cite{jackson1999classical,khosravi2019dirac}. The tensor $T_{ij}$ is composed of three parts :$\nabla_i T_{ij}= \nabla_i (\chi_{ij} + J_{ij} + N_{ij}) $ with    
\begin{subequations}\label{Eq:Terms}
     \begin{equation}\label{Eq:Chirality_&_Helicity}
            \chi_{ij}\! =\Big[ \overbrace{\frac{\hbar}{2}(\psi^\dagger \gamma^5 \psi)}^\text{chirality} + \overbrace{\frac{1}{\mu_0} ( \bm A^\bot\!\cdot\! \bm B)}^\text{helicity}\Big]\delta_{ij} \equiv \chi ~ \delta_{ij},
        \end{equation}
     \begin{equation}\label{Eq:TAM_Currents}
            \begin{split}
                J_{ij} = & \overbrace{\mathcal{R} \left\{\bar{\psi}\gamma_i \left(\bm r\times \bm p_\parallel \right)_j \psi\right\}}^\text{OAM current tensor (Dirac)} - \overbrace{\frac{1}{\mu_0} (A_i^\bot B_j )}^\text{helicity current tensor (Maxwell)} \\
               - & \underbrace{\frac{1}{\mu_0} \left[\varepsilon_{ikl} B_k (\bm r \times \bm\nabla)_j A_l^\bot \right] - \epsilon_0 A_i^\bot \left(\bm r \times \frac{\partial \bm E^\parallel}{\partial t}\right)_j}_\text{OAM current tensor (Maxwell)} ,
            \end{split}
     \end{equation}
        \begin{equation}\label{Eq:EM_Lagrangian}
           \nabla_i N_{ij}= (\bm r \times \nabla) \mathfrak{N}_\text{em}, \quad  \mathfrak{N}_\text{em} =  \underbrace{\frac{\epsilon_0}{2} \bm E^\bot \cdot \bm E^\bot - \frac{1}{2\mu_0}\bm B \cdot \bm B}_\text{free photon Lagrangian} + \epsilon_0 \frac{\partial \bm E^\parallel}{\partial t}\cdot \bm A^\bot.
        \end{equation}
\end{subequations}
where $\gamma^5$ is the chirality operator in Dirac equation (Appendix).

The term $\chi$ in Eq.~(\ref{Eq:Chirality_&_Helicity}) describes the chirality/helicity of the fields. The first term, $\hbar (\psi^\dagger \gamma^5 \psi)/2$, is the density of expectation value (in the classical sense of quantum mechanics) of the chirality operator of the Dirac field. The operator $\gamma^5$ is widely used to find the projections of fermionic fields into left-handed and right-handed chiral states \cite{greiner1996gauge}. Therefore, the expecation value $\psi^\dagger \gamma^5 \psi$ represents the chirality of the Dirac field; a negative (postivie) value means a left-handed (right-handed) chiral state. The second term in Eq.~(\ref{Eq:Chirality_&_Helicity}), on the other hand, is widely known as the helicity of EM field \cite{woltjer1958theorem,coles2012chirality,leeder2015point}. Interestingly, Eq.~(\ref{Eq:Chirality_&_Helicity}) shows that the chirality of the fermionic field is identified with the helicity of the electromagnetic field and the gradient of these quantities enter the continuity equation. 

Note that the chirality of EM field defined in Ref.~\cite{lipkin1964existence} does not appear in the continuity equation of angular momentum. This shows that the helicity is more fundamental for interacting systems. It is also important to note that helicity can be defined for the Dirac field. Being defined as the projection of spin on the momentum of the field, it is fundamentally different from chirality as it is momentum dependent, and unlike chirality, it is not Lorentz covariant. Interestingly, however, for the massless Dirac field, the definition of chirality and helicity become identical \cite{meredith2018helicity}. This points to a similar property in the massless EM fields of the Maxwell's equations.

In the next three sections we define the new terms encountered in the conservation equation. 

\subsection{Helicity current density tensor}
The second terms in Eq.~(\ref{Eq:TAM_Currents}) is reminiscent of the EM helicity in Eq.~(\ref{Eq:Chirality_&_Helicity}) with the difference that this quantity is a tensor, while helicity is a scalar. In fact, helicity density in Eq.~(\ref{Eq:Chirality_&_Helicity}) is equal to the trace of $\frac{1}{\mu_0} A_i^\bot B_j$. We thus name this tensor quantity \textit{helicity current density tensor} since it transforms as a helicity. Furthermore, the divergence of the helicity current tensor  enters the conservation of angular momentum equation meaning that it serves as a current between different forms of Dirac or Maxwell angular momentum. 

One important fact regarding this term is that it has no analog in the Dirac field representation, as one might expect. The equivalent expression for helicity current tensor, in the Dirac field, is normally considered to be spin current represented by a tensor involving $\gamma$ matrices and spin operators. However, our derivation shows that this term vanishes due to the spinor nature of the Dirac fields (Appendix). This, of course, does not put the numerous works on spin currents in condensed matter physics into question \cite{mccreary2012magnetic,zhao2006coherence,Wolf1488,Sharma2005how}, since spin currents are generated by the collective motion of electrons through spatially separating electrons with opposite spins in a quantum cavity \cite{watson2003experimental} or using ferromagnets with strong electron-electron interactions \cite{kashiwaya1999spin}. The relativistic treatment of the Dirac equation in this paper, however, only incorporates the wave function of a single electron in vacuum and thus cannot account for the observed electronic spin currents. This shows that a single electron in vacuum does not exhibit spin current and that  many-body Dirac equation should be considered for a relativistic account of electronic spin currents \cite{CRATER1983two,relativistic1951salpeter}. 

\subsection{Orbital angular momentum current density tensor}

Equation~(\ref{Eq:TAM_Currents}) specifies the angular momentum current tensors for the fields. This current tensor is specified by two directions: direction of angular momentum and direction of the current. For instance, in the first term of Eq.~(\ref{Eq:TAM_Currents}), the direction of propagation of the current is specified by the gamma matrices $\gamma^i$, while the direction of the angular momentum is specified by the operator $(\bm r\times \bm p_\parallel)$. Thus we call this term \textit{OAM current density tensor} of Dirac field. Similarly, the third term in Eq.~(\ref{Eq:TAM_Currents}) is the OAM current tensor density of the EM field. The last term is the angular momentum current tensor due to longitudinal component of electric field and we incorporate that into the OAM current tensor of the EM field. Equation~(\ref{Eq:EM_Lagrangian}) shows that the electromagnetic Lagrangian also contributes to the continuity Eq.~(\ref{Eq:Conservation_Law_02}). 

S. Barnett introduced the angular momentum carrying terms in the source-free electromagnetism \cite{barnett2001optical} as quantities that behave as the currents in the conservation equations written for angular momentum of light. The terms helicity and OAM current tensors introduced here for the electromagnetic field have a similar tensorial nature and extend to the QED interactions. These quantities are gauge-invariant and connect to similar terms coming from the Dirac fields. 

Note that the expression for the angular momentum due to the electromagnetic field in Eq.~(\ref{Eq:TAM}) originally has contribution from the longitudinal electric field, $\bm E^\parallel$ . This is the contribution due to the free charges ($\nabla\cdot\bm E^\parallel = \rho/\epsilon_0$) which decays as a function of $1/r^2$ outside of the region of charges \cite{jackson1999classical}(Appendix). We have shown that the contribution from this component, namely the term $\bm E^\parallel\times \bm A^\bot + E_i^\parallel (\bm r \times \nabla)A_i^\bot$, can be written as the divergence of some quantities involving $\bm E^\parallel$ (Appendix). Therefore, when integrated over the entire space, these terms vanish and thus the total global angular momentum of the interacting field does not involve longitudinal electric field components.
Therefore, we have shown that the angular momentum density of the EM field only depends on the transverse component of the electric field $\bm E^\bot$ since it is only these terms that give nonzero contribution to the total angular momentum.

It should be emphasized that, when working with global quantities (integrated over entire space), the contribution from the longitudinal electric field as well as the contribution from $T_{ij}$ in Eq.~(\ref{Eq:Terms}) become surface terms and vanish assuming that the field quantities are zero on the surface of the integration volume. In such cases, the global angular momentum can be written in equivalent forms which are different only by the divergence of a function of the fields. In the local case, however, all of the terms in Eq.~(\ref{Eq:Terms}) are observable quantities and each have a different interpretation.

\setlength{\extrarowheight}{10mm}
\begin{figure*}[t!]
    \centering
    \setlength\arrayrulewidth{1pt}
    \begin{tabular}{cc|c|c|c|c}
        \vspace{-11.5mm} & &  \multicolumn{4}{l}{\hspace{3mm}\raisebox{12mm}{\footnotesize{time-derivative}} \hspace{2.5mm} \raisebox{11mm}{\large$+$} \hspace{2.5mm} \raisebox{12mm}{\footnotesize{divergence of}} \hspace{2mm} \raisebox{11mm}{\large$+$} \hspace{3.5mm}  \raisebox{12mm}{\footnotesize{gradient of}} \hspace{4.5mm} \raisebox{11mm}{\large$=$} \hspace{3.5mm}  \raisebox{12mm}{\footnotesize{spin-orbit}}} \\
        \vspace{-24mm} & & \multicolumn{4}{l}{\hspace{9mm} \raisebox{24mm}{\footnotesize{of spin}} \hspace{10mm} \raisebox{24mm}{\footnotesize{helicity current tensor}} \hspace{8mm} \raisebox{24mm}{\footnotesize{helicity}} \hspace{20mm}  \raisebox{24mm}{\footnotesize{torque}}} \\ \hline
        \vspace{-10mm} & \Large $\hat{\rho}$ & \hspace{1.5mm} \raisebox{-0.5 \totalheight}{\includegraphics[width = 0.16\textwidth]{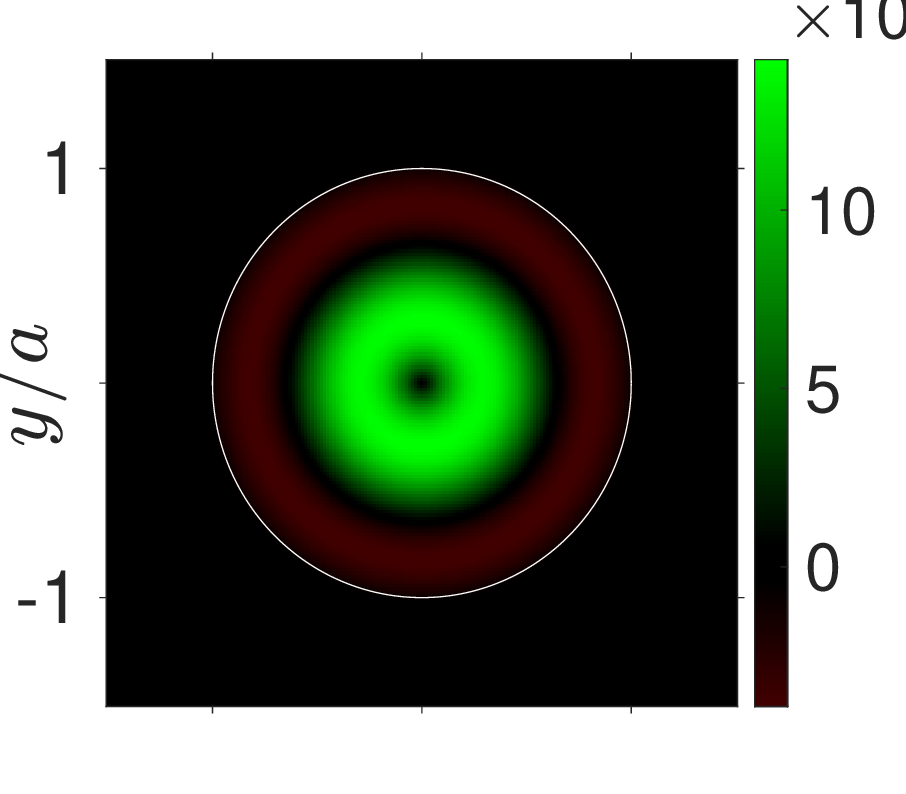}} \hspace{1.5mm} & \hspace{1.5mm}\raisebox{-0.5\totalheight}{\includegraphics[width = 0.16\textwidth]{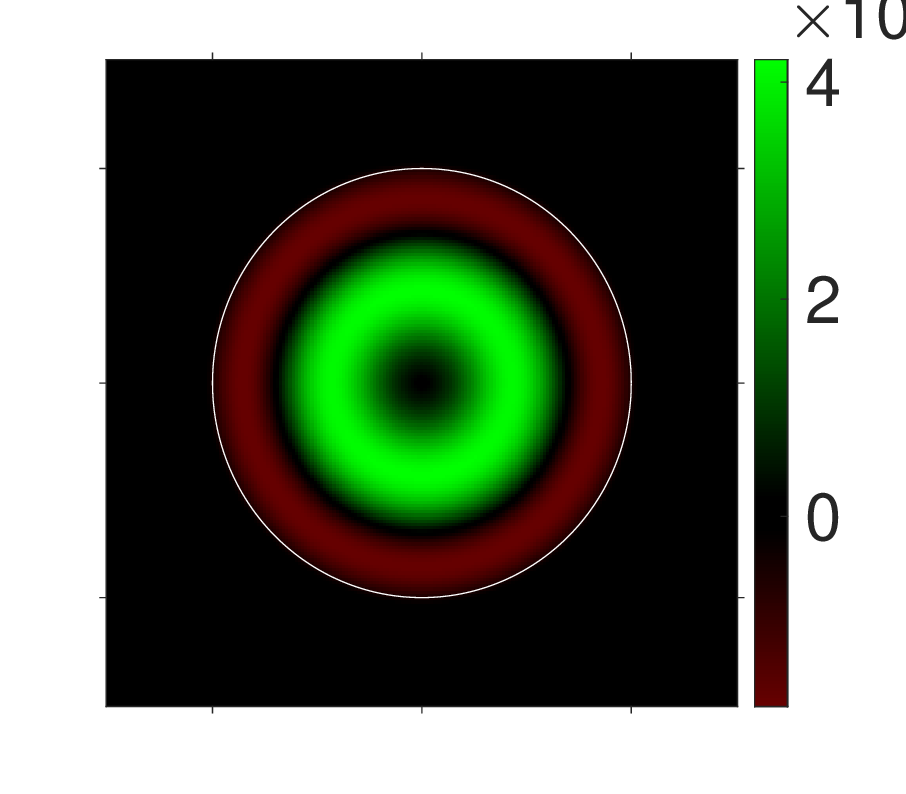}} \hspace{1.5mm}  & \hspace{1.5mm}\raisebox{-0.5\totalheight}{\includegraphics[width = 0.16\textwidth]{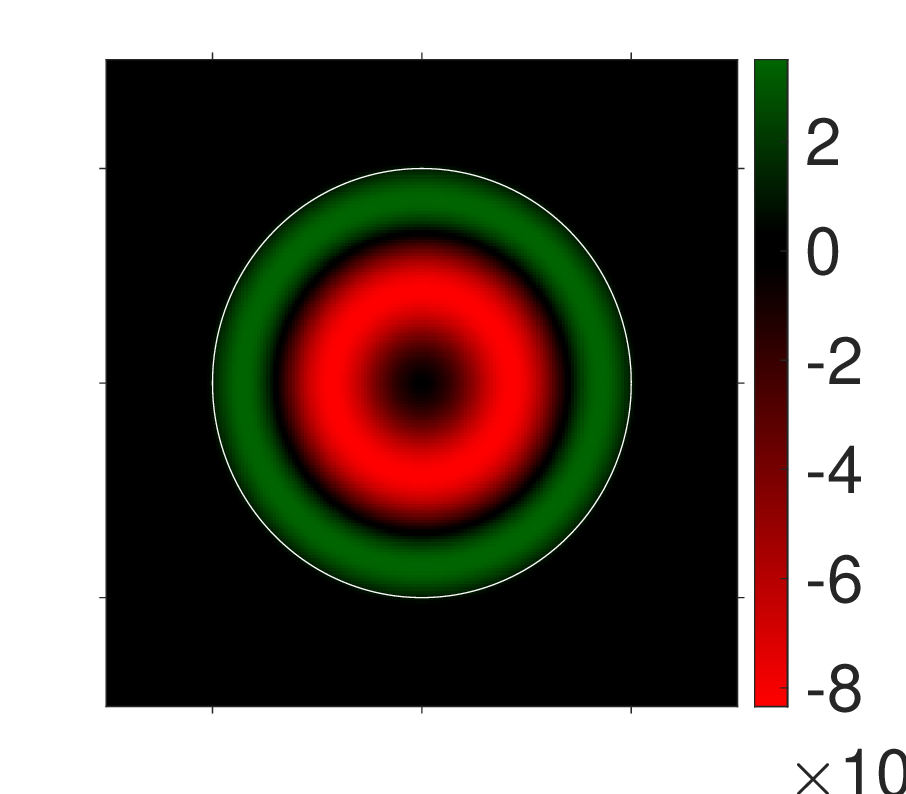}} \hspace{1.5mm}  & \hspace{1.5mm}\raisebox{-0.5\totalheight}{\includegraphics[width = 0.16\textwidth]{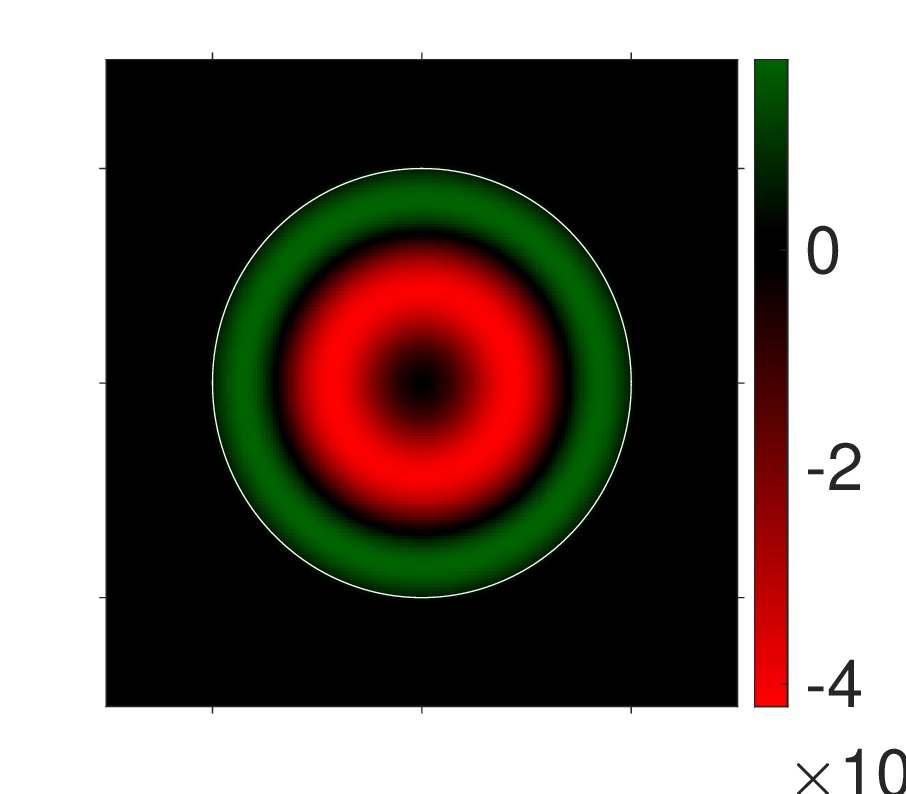}} \hspace{1.5mm} \\
        \vspace{-10mm} \hspace{1.5mm}\rotatebox[origin=c]{90}{Dual-Mode Optical Fiber} & \Large $\hat{\phi}$ &   \hspace{1.5mm}\raisebox{-0.5\totalheight}{\includegraphics[width = 0.16\textwidth]{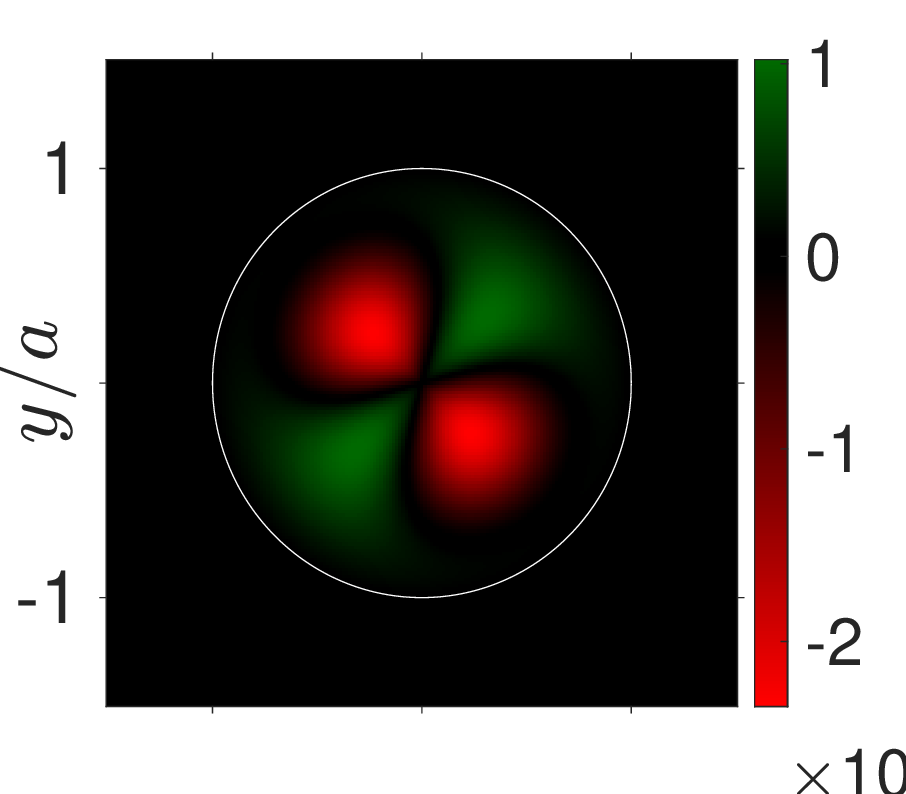}} \hspace{1.5mm} & \hspace{1.5mm}\raisebox{-0.5\totalheight}{\includegraphics[width = 0.16\textwidth]{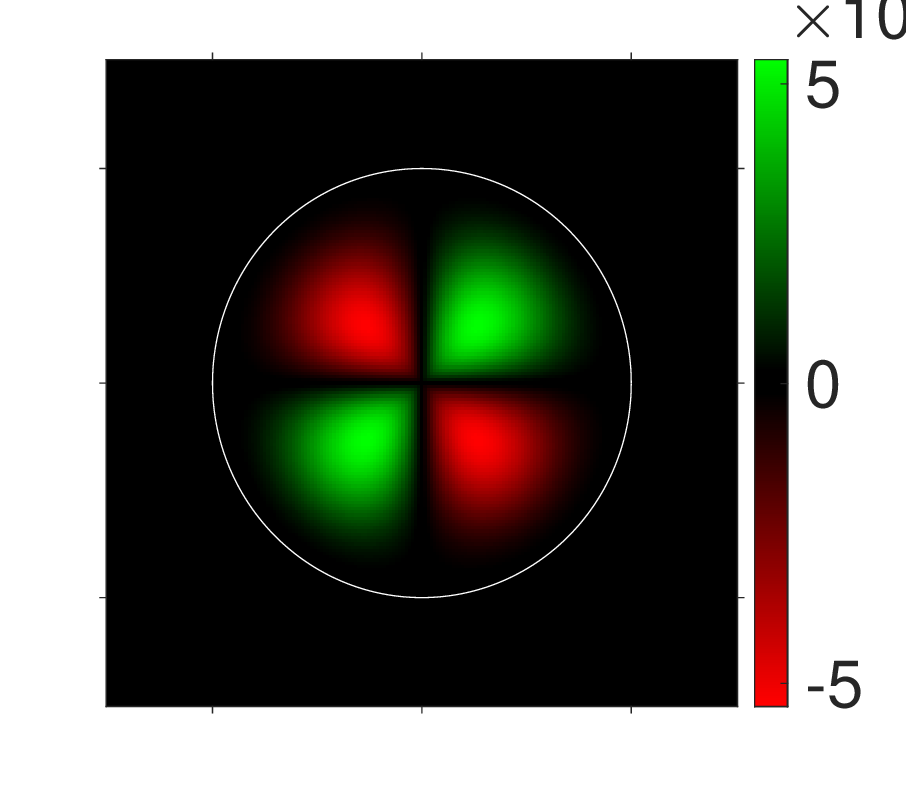}} \hspace{1.5mm} & \hspace{1.5mm}\raisebox{-0.5\totalheight}{\includegraphics[width = 0.16\textwidth]{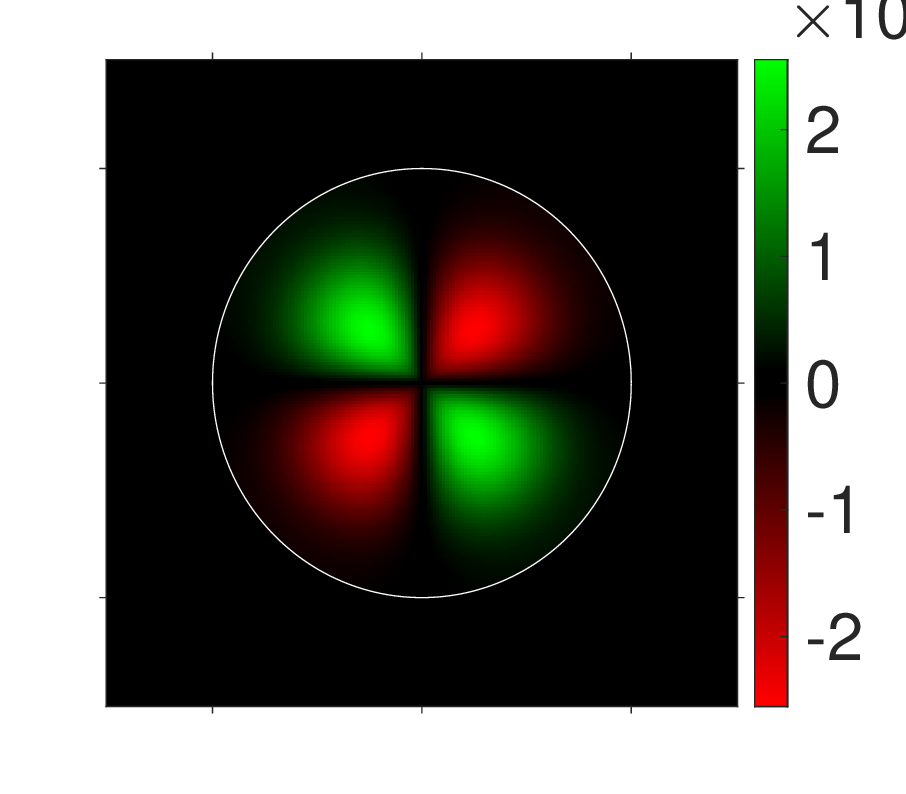}} \hspace{1.5mm} & \hspace{1.5mm}\raisebox{-0.5\totalheight}{\includegraphics[width = 0.16\textwidth]{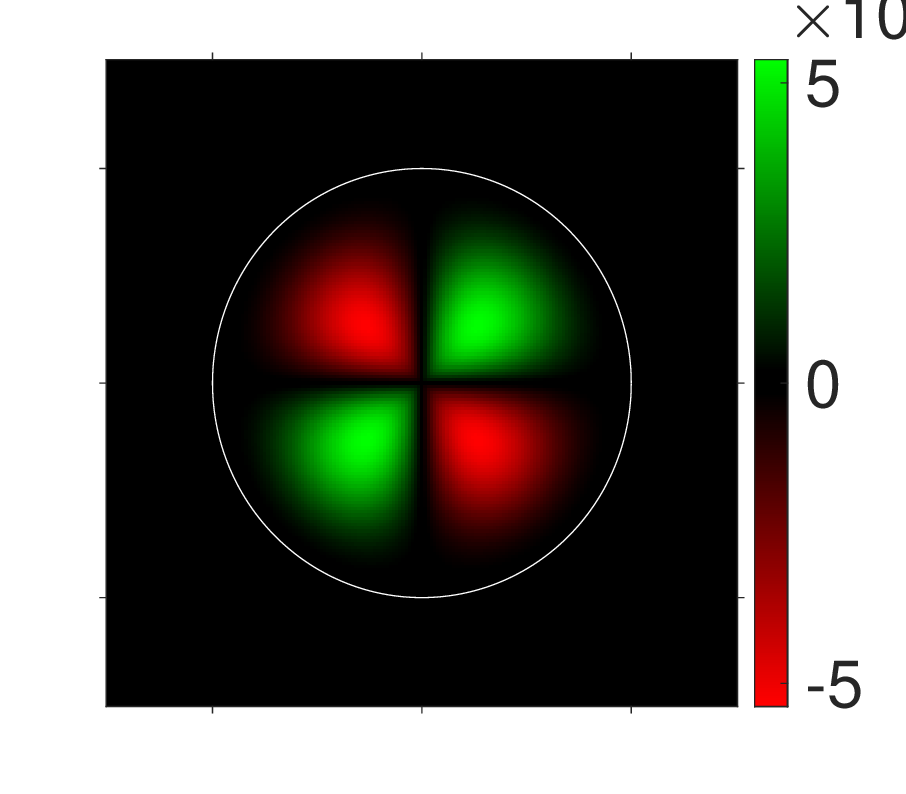}} \hspace{1.5mm} \\ 
        & \Large $\hat{z}$ &   \hspace{1.5mm}\raisebox{-0.5\totalheight}{\includegraphics[width = 0.16\textwidth]{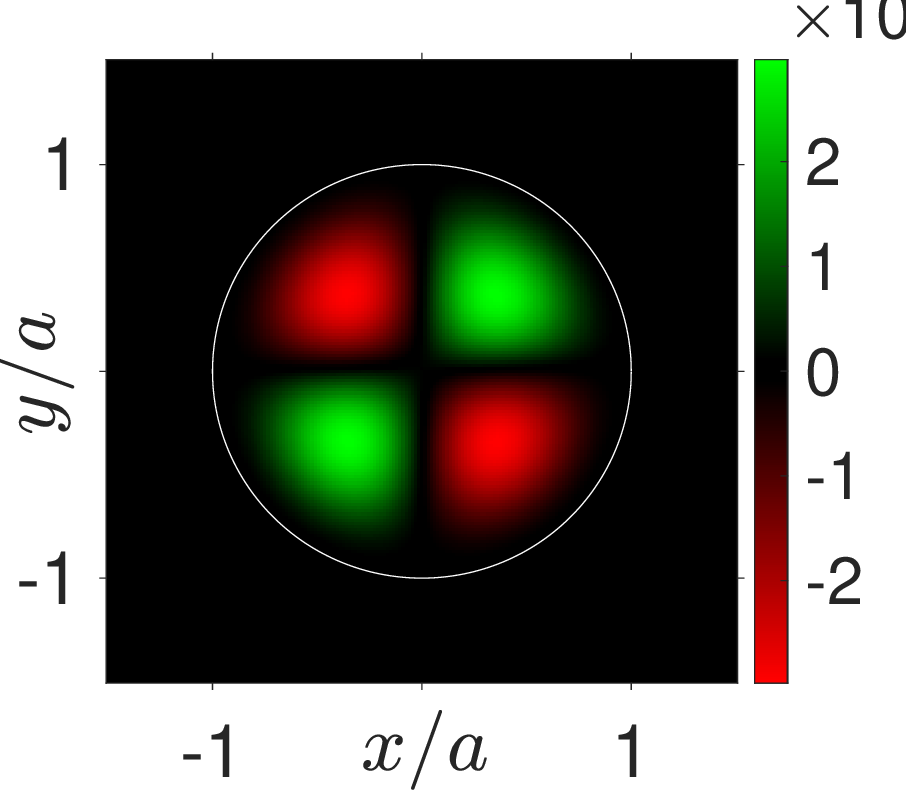}}  \hspace{1.5mm} & \hspace{1.5mm}\raisebox{-0.5\totalheight}{\includegraphics[width = 0.16\textwidth]{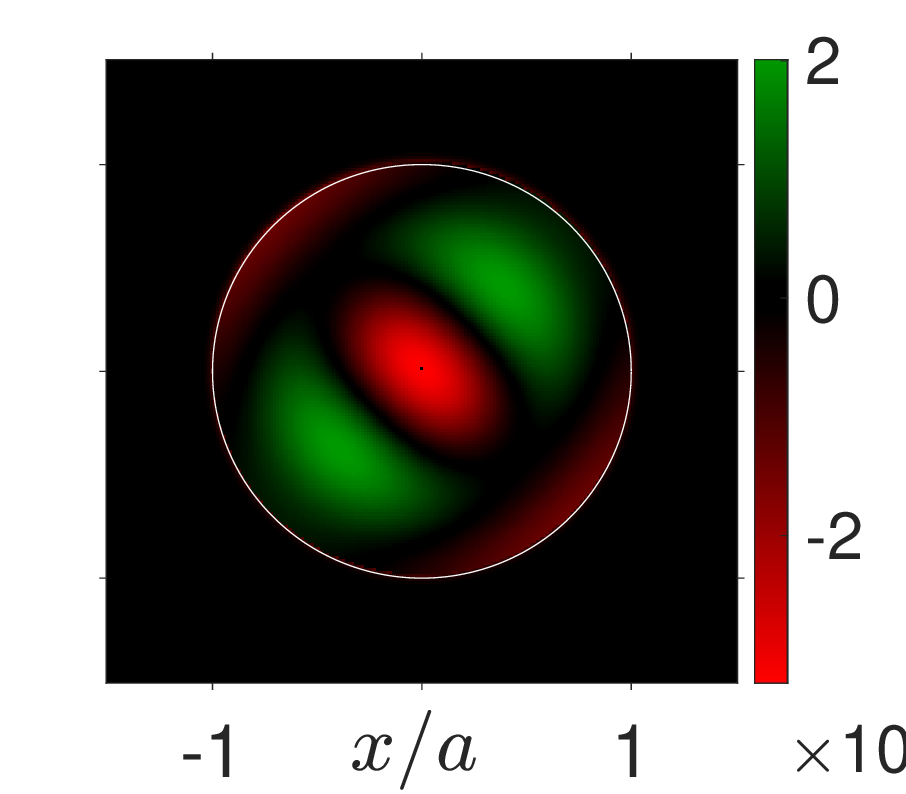}} \hspace{1.5mm} & \hspace{1.5mm}\raisebox{-0.5\totalheight}{\includegraphics[width = 0.16\textwidth]{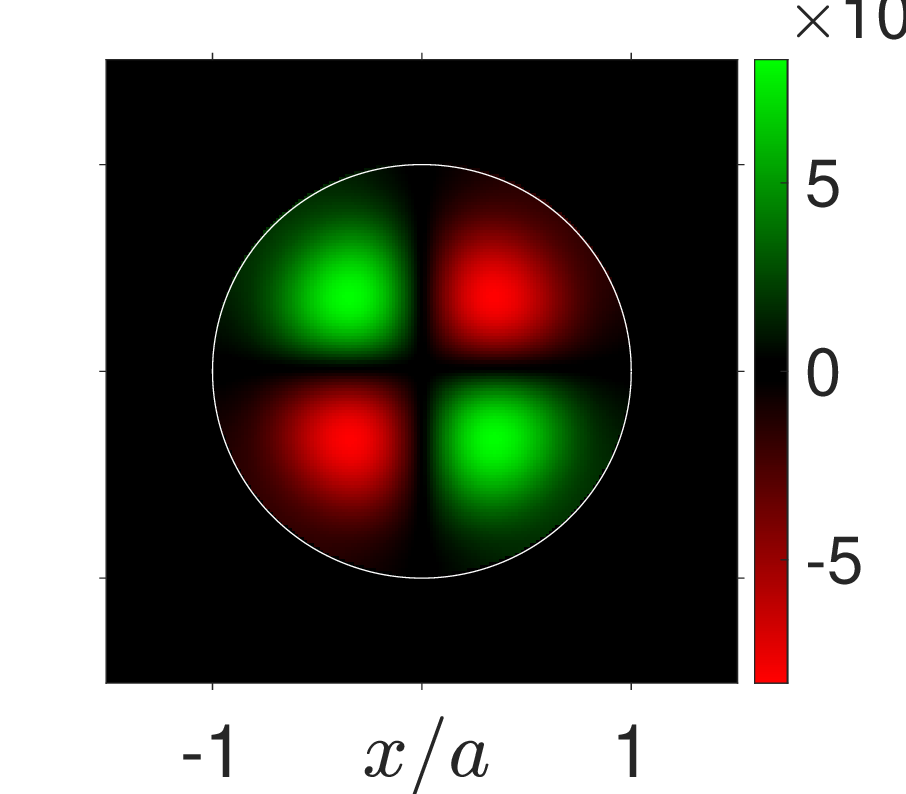}} \hspace{1.5mm} & \hspace{1.5mm}\raisebox{-0.5\totalheight}{\includegraphics[width = 0.16\textwidth]{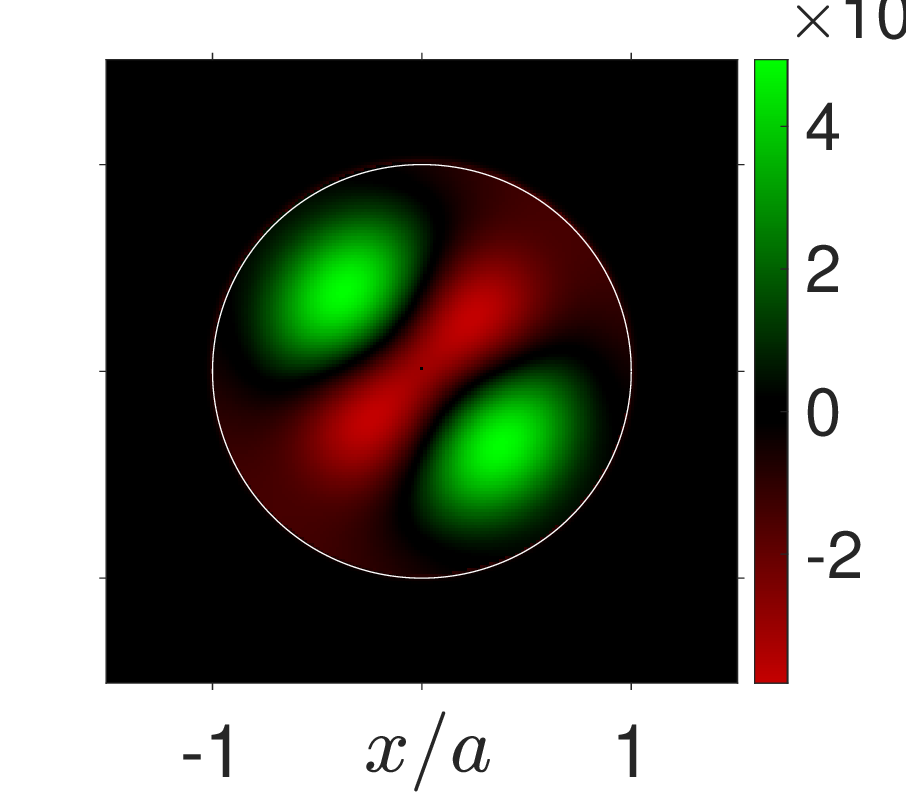}} \hspace{1.5mm}  \\ \hline
    \end{tabular}
    \caption{Local conservation law for the spin [Eq.~(\ref{Eq:Div_Spin_Current_EM})] in a dual-mode optical fiber. The individual terms in Eq.~(\ref{Eq:Div_Spin_Current_EM}) are plotted on the first three columns from left: The first column is the time-derivative of spin ($\frac{\partial}{\partial t} (\bm E^\bot\times A^\bot)$), the second column is the divergence of helicity current tensor ($-\nabla_i(A_i^\bot \bm B)/\mu_0$) , and the third column is the gradient of the helicity density ($\nabla(\bm A^\bot \cdot \bm B)/\mu_0$) . The fourth column is the EM spin-orbit torque given by Eq.~(\ref{Eq:SO_EM}) (in this case $\tau_\text{em}=(\bm B \cdot \nabla)\bm A^\bot /\mu_0 $). The three rows show the local value of each vector along the three axes of optical fiber problem: $\hat{\rho}$ radial direction, $\hat{\phi}$ azimuthal direction, and $\hat{z}$ axis of the fiber. Note that adding the first three column on each row together gives the last column $\bm \tau_\text{em}$; thus confirming Eq.~(\ref{Eq:Div_Spin_Current_EM}). The results are for an optical fiber of radius $50~\mu$m with the two modes at the wavelengths $4.3~\mu$m and $4.29~\mu$m.}
    \label{Fig:Terms_OpticalFiber}
\end{figure*}

\subsection{Spin-orbit torque} 
We now study the spin-orbital angular momentum exchange and the angular momentum transfer between the Dirac and EM fields. The continuity equation in Eqs.~(\ref{Eq:Conservation_Law_01}) and (\ref{Eq:Conservation_Law_02}) only give the dynamics of the total angular momentum density. To obtain insight into the detailed interaction between spin and OAM of the Dirac and EM fields, we separately write the four-divergence of the gauge-independent spin and OAM tensors of the Dirac ($\text{S}^{ij,\lambda}_\text{D}$ and $\text{L}^{ij,\lambda}_\text{D}$) as well as the EM fields ($\text{S}^{ij,\lambda}_\text{em}$ and $\text{L}^{ij,\lambda}_\text{em}$). By incorporating the Dirac and Maxwell's equations into the gauge-invariant angular momentum densities in Eqs.~(\ref{Eq:TAM}) and (\ref{Eq:Terms}), we obtain their detailed conservation equations  (see Appendix),
\begin{subequations}\label{Eq:Div_SO}
    \begin{equation}\label{Eq:Div_Spin_Dirac}
        \partial_\lambda \text{S}_\text{D}^{ij,\lambda} = \varepsilon_{ijk}\left[\bm \tau_\text{D} + \bm j_c \times \bm A^\bot\right]_k,
    \end{equation}    
    \begin{equation}\label{Eq:Div_OAM_Dirac}
       \!\!\! \partial_\lambda \text{L}_\text{D}^{ij,\lambda}\! = \!\varepsilon_{ijk} \!\left[ -\bm \tau_\text{D}\! + \! \rho(\bm r\times \bm E^\parallel) \!+\! j_{c,l}(\bm r\times \bm\nabla) A_l^\bot \right]_k \!\!,
    \end{equation}
    \begin{equation}\label{Eq:Div_Spin_EM}
        \partial_\lambda \text{S}_\text{em}^{ij,\lambda} = \varepsilon_{ijk} \left[ \bm \tau_\text{em} - \bm j_c \times \bm A^\bot \right]_k,
    \end{equation}
    \begin{equation}\label{Eq:Div_OAM_EM}
       \!\!\! \partial_\lambda \text{L}_\text{em}^{ij,\lambda} \!=\! \varepsilon_{ijk}\! \left[-\bm \tau_\text{em}\! -\! \rho(\bm r\times \bm E^\parallel)\! - \!j_{c,l} (\bm r\times \bm\nabla) A^\bot_l \right]_k \!\!,
    \end{equation}
\end{subequations}
where we have defined
\begin{equation}\label{Eq:SO_Dirac}
    \bm \tau_\text{D} = - c \mathcal{R}\left\{ \bar{\psi}\left(\bm \gamma \times\bm p_\parallel\right)\psi \right\}
\end{equation}
and 
\begin{equation}\label{Eq:SO_EM}
    \bm \tau_\text{em} = \frac{1}{\mu_0}(\bm B \cdot \nabla)\bm A^\bot - \epsilon_0 \left(\frac{\partial \bm E^\parallel}{\partial t} \times \bm A^\bot \right)
\end{equation}
as the \textit{Dirac} and \textit{Maxwell} \textit{spin-orbit torque}, respectively, since it gives the amount of torque exerted on the spin from the OAM of the fields and vice versa. This nomenclature is further motivated by the resemblance of the Dirac spin-orbit torque (Eq.~(\ref{Eq:SO_Dirac})) of the Rashba spin-orbit coupling Hamiltonian \cite{manchon2015new}. The direct connection between these terms, however, is out of scope of this article and is the focus of a future work. Note that $\bm{\gamma} = \gamma^1 \hat{x} + \gamma^2 \hat{y} + \gamma^3 \hat{z}$ and $\bm j_c = ec \bar{\psi}\bm \gamma \psi$ is the electric charge current density in Eqs.~(\ref{Eq:Div_SO}) and (\ref{Eq:SO_Dirac}). 

Equation (\ref{Eq:Div_SO}) clearly shows that the spin-orbit torques contribute to the spin-OAM exchange in both Dirac and EM fields. Moreover, it is evident from Eqs.~(\ref{Eq:Div_Spin_Dirac}), (\ref{Eq:Div_OAM_Dirac}), (\ref{Eq:Div_Spin_EM}), and (\ref{Eq:Div_OAM_EM}) that the charge-field coupling terms, $\bm j_c\times \bm A^\bot$, $j_{c,l}(\bm r \times \nabla)A_l^\bot$, and $\rho(\bm r\times \bm E^\parallel)$, are responsible for the angular momentum transfer between the Dirac and EM fields \cite{van1994commutation}. In fact, the first terms gives rise to an optical torque exerted on dipoles due to a circularly polarized optical field \cite{nieto2015optical}. Our results are significantly different from scalar continuity equations of EM helicity in Refs.~\cite{crimin2019optical} and~\cite{bliokh2013dual}, which derive a scalar conservation law for the dual-symmetric expressions of spin and helicity of EM field. The helicity continuity equation obtained from the free-space Maxwell equations can not characterize the spin-OAM exchange and specifically the angular momentum transfer between Dirac and EM fields.

\section{Source-free problems}
We now show the importance of these continuity equations by demonstrating the spin-OAM exchange via the spin-orbit torque for propagating EM fields. We evaluate the terms in $\partial_\lambda \text{S}^{ij,\lambda}_\text{em}$ and the EM spin-orbit torque $\bm \tau_\text{em}$ (Eq.~(\ref{Eq:SO_EM})) for two simple EM problems in source-free regions. Similar spin-orbit signature can also be observed in a cylindrical geometry for the Dirac fields \cite{khosravi2019dirac}. However, a thorough study of each individual term in Eq.~(\ref{Eq:Terms}) for Dirac and EM fields is not the purpose of this paper. 

The general form of $\partial_\lambda \text{S}^{ij,\lambda}_\text{em}$, in a source-free regions, is 
\begin{equation}\label{Eq:Div_Spin_Current_EM}
    \epsilon \frac{\partial}{\partial t}(\bm E^\bot \times \bm A^\bot\!) - \frac{1}{\mu_0}\nabla_i (A_i^\bot\! \bm B)\! +\! \frac{1}{\mu_0}\! \nabla (\!\bm A^\bot\! \cdot\! \bm B)\!= \bm \tau_{em}.\!
\end{equation}
Here, the extra term on the right hand side results from the coupling to OAM. The spin-orbit torque in the source-free case reduces to $\bm \tau_{em}\! =\!  (\bm B \cdot\nabla)\bm A^\bot/\mu_0$. We emphasize that different from free-space case~\cite{crimin2019optical,bliokh2013dual}, the near-field spin-OAM exchange can still exist in the presence of sources.


\subsection{dual-mode optical fiber}
We first consider a dual-mode optical fiber, which is placed along the $z$ axis with radius $a$. The two modes have the same propagation constant, $\beta$, and different frequencies of $\omega_1$ and $\omega_2$. Figure~\ref{Fig:Terms_OpticalFiber} shows the four terms in Eq.~(\ref{Eq:Div_Spin_Current_EM}). The three rows show the components of these quantities along the radius $\hat{\rho}$, azimuthal $\hat{\phi}$, and $z$ axis of the fiber. Note that for all three rows, the sum of the first three terms is equal to $\bm \tau_\text{em}$; thus satisfying Eq.~(\ref{Eq:Div_Spin_Current_EM}). These figures show that, for a dual-mode optical fiber, the helicity current current tensor, helicity density, spin density, as well as the spin-orbit torque are all non-zero and can play a role in a local interaction between optical modes and atomic sources. Note that all three components become zero as the fiber radius $a\to \infty$, which confirms that the spin-orbit torque $\bm \tau_\text{em}$ becomes zero for plane wave solutions (Appendix). 

\begin{figure}[h!]
    \centering
    \subfloat[\label{Fig:SpinTerms_vs_z}]{
        \includegraphics[width=0.6\linewidth]{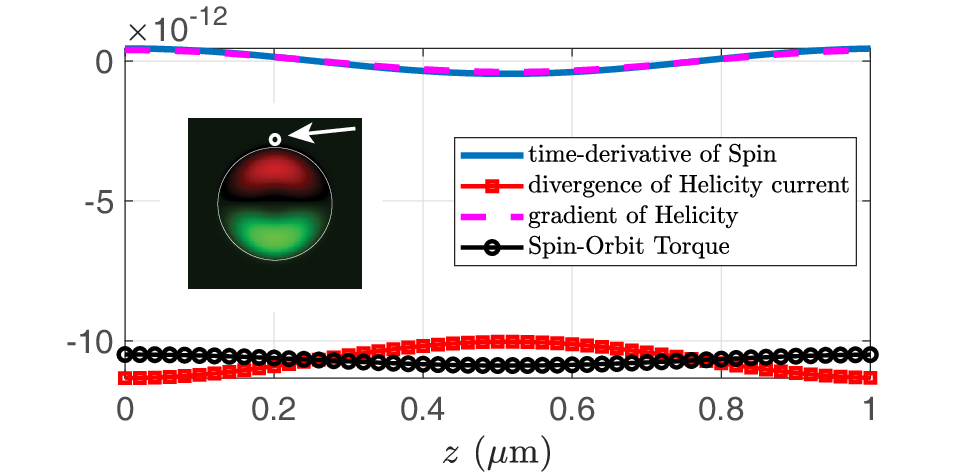}}\par \vspace{-4mm}
        \subfloat[\label{Fig:SpinTerms_vs_t}]{
        \includegraphics[width=0.6\linewidth]{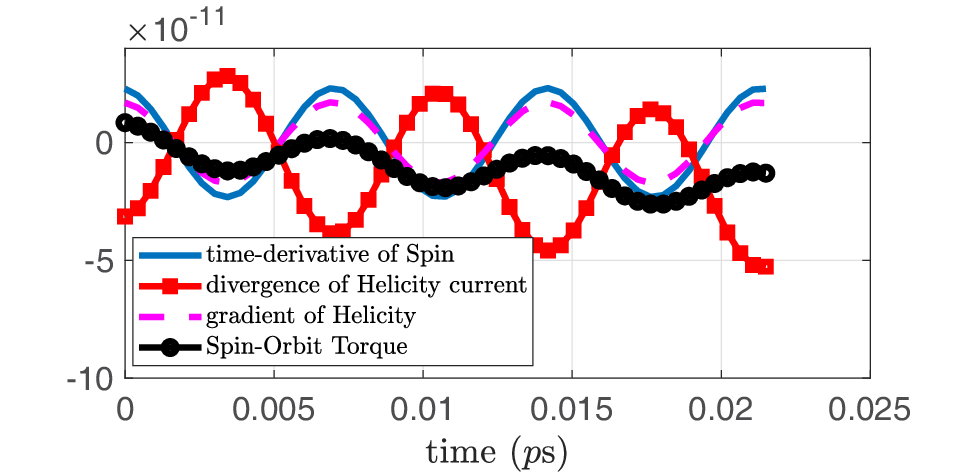}}
    \caption{Dynamics of the terms in Eq.~(\ref{Eq:Div_Spin_Current_EM}) versus (a) $z$ and (b) time. The plots only show the $\hat{z}$ component of each term. These plots show that the spin-orbit torque ($\bm\tau_\text{em}$) is equal to the sum of the other terms. The inset in panel (a) shows the location where the terms are evaluated for both of the figures.}
    \label{Fig:SO-Dynamics}
\end{figure}

Figure~\ref{Fig:SO-Dynamics} shows the dynamics of the time-derivative of spin, divergence of helicity current density, and gradient of helicity (the terms in Eq.~(\ref{Eq:Div_Spin_Current_EM})) as a function of $z$ (Fig.~(\ref{Fig:SpinTerms_vs_z})) and time (Fig.~(\ref{Fig:SpinTerms_vs_t})). The curves show the value of these terms at a point $0.05~a$ from the fiber as shown in the insets. The black curve with the circle markers show the EM spin-orbit torque, $\bm \tau_\text{em}$, which is essentially the torque that is transferred to the OAM of the fields. Note that the sum of other curves equals the black curve $\bm \tau_\text{em}$; satisfying the continuity Eq.~(\ref{Eq:Div_Spin_Current_EM}). 

\subsection{double plane wave interference}
The second example is the interference of
two circularly polarized plane waves at two different frequencies. For the plane wave solutions, the current term $\nabla_i (A_i^\bot \bm B)$ and the spin-orbit torque~$\bm \tau_{em}$ vanish (Appendix). Therefore, we arrive at a simpler conservation law for the second problem, 
\begin{equation}\label{Eq:Conservation_Law_PlaneWaves}
     \frac{\partial}{\partial t} \left(\epsilon_0\bm E^\bot \times \bm A^\bot \right) + c   \bm\nabla \left(\frac{1}{\mu_0 c}\bm A^\bot \cdot \bm B\right) = 0,
\end{equation}
which shows the spin density propagating in space with light speed $c$. For circularly polarized plane waves propagating along $z$ direction, the electric field is written as $\bm E = \mathcal{R}\{ \bm{\mathcal{E}}_1 e^{-i(\omega_1 t - k_1 z)} + \bm{\mathcal{ E}}_2 e^{-i(\omega_2 t - k_2 z)} \}$, where $\bm{\mathcal{E}}_i = \mathcal{E}_i(\hat{x} + i \hat{y})/\sqrt{2}$ are the complex electric field amplitudes of the two modes with frequencies $\omega_i/c = k_i $. For these fields, we find (Appendix)
\begin{equation}
    \begin{split}
        \frac{1}{\mu_0} & \nabla(\bm A^\bot \cdot \bm B ) = - \epsilon_0 \frac{\partial}{\partial t} (\bm E \times \bm A^\bot) \\
        = & \epsilon_0 \frac{\omega_1^2 - \omega_2^2}{2\omega_1 \omega_2} \mathcal{I}\left\{\mathcal{E}_1 \mathcal{E}_2^* e^{-i[(\omega_1 - \omega_2)t - (k_1 - k_2)z]}\right\},
    \end{split}
\end{equation}
hence satisfying the conservation Eq.~(\ref{Eq:Conservation_Law_PlaneWaves}). This clearly shows that the change in time of the spin is compensated by the gradient of the helicity of the EM field.

\section{Experimental perspective}

The interaction between the electronic spin and optical angular momentum has been widely used in devices such as optical circulators, insulators, electro-optic modulators, and Faraday rotators. In these devices, conservation of angular momentum combined with the field properties of the electromagnetic (EM) radiation gives rise to non-reciprocal effects that can be manipulated for a wide variety of applications. These devices incorporate the interaction between EM radiation and the magnetic materials in the far-field regions where a conservation law based on the total electronic and photonic angular momenta suffices to explain the underlying physical phenomena \cite{malinowski2008control}.

Due to the possibility of optical control of the quantum spin states, the dynamics of angular momentum between the electronic and optical fields has recently gained attention for the systems interacting in the near-field region. Coupling to the local spin of EM field has been observed for cold atoms in the vicinity of an optical fiber \cite{mitsch2014quantum}, magnons interacting with spherical whispering gallery modes (WGMs) \cite{schmiegelow2016transfer}, quantum dots in photonic crystals \cite{young2015polarization}, as well as quantum sources coupling to other waveguide systems \cite{sollner2015deterministic,barik2019chiral,tang2010optical}. Non-classical spin texture and non-local photonic spin noise has been shown in quantum structured light~\cite{Das22024arxiv,Yang2021nonclassical}, which can be measured in experiment via nano-scale quantum sensors for photonic spin density~\cite{Kalhor2021quantum}. In such systems, since the interaction between the source and the EM field occurs in the near-field rather than the far-field regions, a local approach to the governing dynamical equations of angular momentum becomes important. These studies reveal the new insight that the local interaction with classical electromagnetic field has to offer. 

With recent advances in the cold atom and quantum dot communities, local interactions between atomic sources and optical fields are gaining more attention due to the emergent new phenomena. These experiments emphasize the need for a local conservation law describing the dynamics of angular momentum in light-matter interactions.
 
We applied the Noether's theorem to the Dirac Lagrangian interacting with the EM field to find the local conservation laws of angular momentum for the most general electrodynamics problem. The results developed here can be applied to near-field as well as far-field to study the transformation of angular momentum between different fields in these regions. Our results show that, in consideration of local conservation laws, other quantities including helicity and OAM current tensors, EM helicity, and electronic chirality should also be considered in addition to the spin and OAM of the EM and Dirac fields. 

Equation~(\ref{Eq:Conservation_Law_02}) holds everywhere in space and time. This shows that the total angular momentum density $\bm M$ is not locally conserved. In other words, $\frac{\partial \bm M}{\partial t}\neq 0$ and the conservation law can only be written after including all the other terms in Eqs.~(\ref{Eq:Conservation_Law_02}) and (\ref{Eq:Terms}). Integrating Eq.~(\ref{Eq:Conservation_Law_02}) over some volume $V$, on the surface of which both Dirac and EM fields become zero, $\psi\to 0$, $\bm E\to 0$, gives the usual global conservation law, $\frac{\partial}{\partial t} \int_V \bm M d^3x = 0$. This states that the integrated values of the spin and OAM densities of electron and EM field over the entire space is a conserved quantity (Fig.~\ref{Fig:Schematic_Old}).

We can also get a simpler continuity equation than Eq.~(\ref{Eq:Conservation_Law_02}) if we limit ourselves to regions outside the Dirac fields. To do so, we take the integral of Eq.~(\ref{Eq:Conservation_Law_02}) in the volume $V'$ on which only the Dirac fields become zero $\psi=0$. Doing so we get the \textit{semi-local} conservation law (Appendix),
\begin{equation}\label{Eq:Conservation_Law_SemiLocal}
    \frac{\partial \Tilde{\bm M}_\text{D}} {\partial t} =  -\left[ \frac{\partial \Tilde{\bm M}_\text{em}}{\partial t} +  \Tilde{\bm J} + \Tilde{\bm h} + \int_{S'} \hat{\bm n} \times (\bm r {\mathfrak{N}}_\text{em})da\right]
\end{equation}
which expresses the time-evolution of the angular momentum of electron $\bm \tilde{\bm M}_\text{D}$ in terms of EM dependent quantities. $\tilde{\bm M}_\text{em}$ is the EM angular momentum in the region of the source, while $\tilde{\bm J}$ and $\tilde{\bm h}$ are the EM angular momentum current tensor (projected onto the normal of the surface) and helicity (multiplied by the normal of the surface), integrated on the surface of the volume surrounding the current charges (Appendix). Also, $\mathfrak{N}_\text{em}$ is given by Eq.~(\ref{Eq:EM_Lagrangian}). Equation (\ref{Eq:Conservation_Law_SemiLocal}) can be used to find the conservation of angular momentum in the near-field of current sources and how is it transferred to the electromagnetic fields.

It is also important to note that the conservation of angular momentum derived here is based on the spatial components of Lorentz transformations (i.e. rotations). Time-space components of the Lorentz transformation can also give conservation equations related to the boost operators in the relativistic equations of motion. These equations provide a new conservation equation for the Dirac-Maxwell fields which can be of interest to applications investigating relativistic behaviour of the particles and their pertinent conserved quantities when interacting. A brief discussion about these conservation equations is given in the Appendix.

The method presented in this paper can be further extended to find the dynamics of magnetization in different materials. The simplest system can be regarded as the interaction between an externally applied EM field and the electrons in a non-magnetic metal. A weak probe signal can then be used to study spin dynamics of the metal. Such a system can be closely modeled as a non-interacting electron gas whose dynamics can be described by the Dirac equation. Although the solutions of the Dirac equation interacting with an externally applied plane wave can be rigorously found \cite{Wolkow1935,redmond1965solution}, these solutions are extremely complicated and only apply to the particular case of free electrons in a plane wave. Our method, circumvents the problem of solving the Dirac Hamiltonian to find the spin dynamics of the electrons, by using Noether's theorem and finding the conservation equation governing angular momentum dynamics of the electrons. By knowing the properties of the externally applied EM field, a simpler equation such as Eq.~(\ref{Eq:Conservation_Law_SemiLocal}) can be used to find the dynamics of angular momentum without the need of electronic wavefunctions. Since in our derivations we make no simplifying assumption on the EM field, these equations can be easily used for interactions that take place in the near-field region and where EM fields cannot be regarded as plane waves.

The Dirac equation can be further extended to model ferromagnetic materials. Dirac-Kohn-Sham (DKS) equation is an extension to the Dirac equation which accounts for the Kohn-Sham potential as well as the spin-polarized part of the exchange correlation potential inside a magnetic material \cite{MacDonald_1979,eschrig1999relativistic}. Corrections from DKS equation can be added to the usual Pauli Hamiltonian to account for the terms in the Landau-Lifshitz-Gilbert (LLG) equation and to add corrections accounting for higher order terms dependent on external and internal parameters \cite{mondal2018}. The method presented in this paper can also be applied to the DKS Hamiltonian to derive the conservation equation for the angular momentum of electrons in magnetic materials.

\section{Acknowledgements}
This work is partially supported by the funding from Army Research Office (W911NF-21-1-0287). L.P.Y is funded by National Key R\&D Program of China (No. 2021YFE0193500). L.P.Y was a post-doctoral scholar and Farhad Khosravi was a visiting scholar at Purdue University when this collaborative research work was performed. 

\section{Data Availability Statement}
All data that support the findings of this study are included within the article (and any supplementary files).

\appendix
\section{\label{Sec:Noether Theorem} Conserved Angular Momentum Tensor}
Symmetrized Dirac Lagrangian, with the minimal coupling term \cite{greiner2013field}, is written as 
\begin{equation}\label{Eq:Dirac_lagrangian}
    \mathfrak{L} = \Bar{\psi}\left[c\gamma^\mu(\frac{i\hbar}{2}\overleftrightarrow{\partial}_\mu - e A_\mu) - mc^2  \right]\psi - \frac{1}{4\mu_0}F_{\mu\nu}F^{\mu\nu}
\end{equation}
where $\overleftrightarrow{\partial} = \overleftarrow{\partial} + \overrightarrow{\partial}$ with $\overleftarrow{\partial}$ and $\overrightarrow{\partial}$ acting only on $\Bar{\psi}$ and $\psi$, respectively, 
\begin{equation}
    F_{\mu\nu} = \partial_\mu A_\nu - \partial_\nu A_\mu = \begin{pmatrix}
        0 & {E_x}/{c} & {E_y}/{c} & {E_z}/{c} \\
        -{E_x}/{c} & 0 & - B_z & B_y \\
        -{E_y}/{c} & B_z & 0 & -B_x \\
        -{E_z}/{c} & - B_y & B_x & 0 
    \end{pmatrix}
\end{equation}
is the electromagnetic tensor, and $\gamma^\mu$ are the Dirac gamma matrices with the property $\{\gamma^\mu,\gamma^\nu \} = \eta^{\mu\nu}$, where
\begin{equation}
    \eta^{\mu\nu} = \begin{pmatrix}
    1 &  0 & 0 & 0 \\
    0 & -1 & 0 & 0 \\
    0 &  0 & -1 & 0 \\
    0 &  0 &  0 & -1
    \end{pmatrix}
\end{equation}
is the Minkowski metric tensor with the signature $(+ - - -)$. We can get the conserved currents related to the rotational symmetry of the Lagrangian, using the Noether's theorem, as \cite{noether1971invariant,greiner2013field}:
\begin{equation}\label{Eq:Noether_currents_01}
    \mathcal{M}^{\mu\nu,\lambda} = \left( -\frac{i}{2}\frac{\partial \mathfrak{L}}{\partial(\partial_\lambda\psi)}\hat{M}_\text{D}^{\mu\nu}\psi \right) + \left( -\frac{i}{2}\Bar{\psi}\hat{M}_\text{D}^{\mu\nu}\frac{\partial\mathfrak{L}}{\partial(\partial_\lambda \Bar{\psi})} \right) + \left(-\frac{i}{2} \frac{\partial \mathfrak{L}}{\partial(\partial_\lambda A^\kappa)}\right) {(\hat{M}_\text{em}^{\mu\nu})^\kappa}_\sigma A^\sigma + \frac{1}{2}\left(\eta^{\lambda\mu}\mathfrak{L}x^\nu - \eta^{\lambda\nu}\mathfrak{L}x^\mu\right)
\end{equation}
where
\begin{equation}
    \hat{M}_\text{D}^{\mu\nu} = \hat{L}^{\mu\nu} + \hat{\Sigma}^{\mu\nu} 
\end{equation}
is the angular momentum operator for the Dirac fields with
\begin{equation}\label{Eq:Spin_Operator_Dirac}
    \hat{\Sigma}^{\mu\nu} = \frac{1}{2}\sigma^{\mu\nu},\quad \sigma^{\mu\nu} = \frac{i}{2}[\gamma^\mu,\gamma^\nu]
\end{equation}
\begin{equation}\label{Eq:OAM_Operator}
    \hat{L}^{\mu\nu} = x^\mu \partial^\nu - x^\nu \partial^\mu,
\end{equation}
and 
\begin{equation}
    {(\hat{M}_\text{em}^{\mu\nu})^\kappa}_\sigma = \hat{L}^{\mu\nu}\delta^\kappa_\sigma + {(\hat{S}^{\mu\nu})^\kappa}_\sigma
\end{equation}
is the angular momentum operator for the electromagnetic fields with $\hat{L}^{\mu\nu}$ given by Eq.~(\ref{Eq:OAM_Operator}), $\delta^\kappa_\sigma$ being the Kronecker delta function, and 
\begin{equation}
    {(\hat{S}^{\mu\nu})^\kappa}_\sigma = i\left(\eta^{\mu\kappa}{\eta^\nu}_\sigma - {\eta^\mu}_\sigma \eta^{\nu\kappa}\right).
\end{equation}
Plugging these equations into Eq.~(\ref{Eq:Noether_currents_01}), we get for the angular momentum currents
\begin{equation}
    \mathcal{M}^{\mu\nu,\lambda} = \mathcal{M}_\text{D}^{\mu\nu,\lambda} + \mathcal{M}_\text{em}^{\mu\nu,\lambda} 
\end{equation}
where
\begin{subequations}\label{Eq:AM_Current_Dirac_01}
 \begin{equation}
     \mathcal{M}_\text{D}^{\mu\nu,\lambda} = \mathcal{S}_\text{D}^{\mu\nu,\lambda} + \mathcal{L}_\text{D}^{\mu\nu,\lambda}
 \end{equation}
 \begin{equation}
     \mathcal{S}_\text{D}^{\mu\nu,\lambda} = \frac{\hbar c}{4}\bar{\psi}\left( \gamma^\lambda \sigma^{\mu\nu} + \sigma^{\mu\nu}\gamma^\lambda \right)\psi
 \end{equation}
 \begin{equation}
     \mathcal{L}_\text{D}^{\mu\nu,\lambda} = \hbar c \mathcal{R}\left\{ \Bar{\psi}\gamma^\lambda (x^\mu\partial^\nu - x^\nu \partial^\mu)\psi \right\} + (\eta^{\lambda\mu}x^\nu - \eta^{\lambda\nu}x^\nu)\mathfrak{L}_\text{D}
 \end{equation}
 \begin{equation}
     \mathfrak{L}_\text{D} = \Bar{\psi}\left[ i\hbar c \frac{1}{2}\gamma^\mu \overleftrightarrow{\partial}_\mu - mc^2 \right]\psi - c e \Bar{\psi}\gamma^\mu\psi A_\mu
 \end{equation}
\end{subequations}
is the contribution due to the Dirac field and
\begin{subequations}\label{Eq:AM_Current_EM_01}
    \begin{equation}
        \mathcal{M}_\text{em}^{\mu\nu,\lambda} = \mathcal{S}_\text{em}^{\mu\nu,\lambda} + \mathcal{L}_\text{em}^{\mu\nu,\lambda}
    \end{equation}
    \begin{equation}
        \mathcal{S}_\text{em}^{\mu\nu,\lambda} = -\frac{1}{\mu_0} \left(F^{\lambda\mu}A^\nu - F^{\lambda\nu}A^\mu\right)
    \end{equation}
    \begin{equation}
        \mathcal{L}_\text{em}^{\mu\nu,\lambda} = -\frac{1}{\mu_0}\left[ {F^\lambda}_\kappa (x^\mu\partial^\nu - x^\nu \partial^\mu)A^\kappa \right] + (\eta^{\lambda\mu}x^\nu - \eta^{\lambda\nu}x^\mu)\mathfrak{L}_\text{em}
    \end{equation}
    \begin{equation}
        \mathfrak{L}_\text{em} = -\frac{1}{4\mu_0}F_{\mu\nu}F^{\mu\nu} = \frac{1}{2\mu_0}\left(\frac{\bm E\cdot\bm E}{c^2} - \bm B\cdot\bm B\right)
    \end{equation}
\end{subequations}
is the contribution due to the electromagnetic field. As a consequence of Noether theorem, the angular momentum tensor $\mathcal{M}^{\mu\nu,\lambda}$ is conserved. In other words,
\begin{equation}\label{Eq:Conservation_Law_01}
    \partial_\lambda \mathcal{M}^{\mu\nu,\lambda} = 0.
\end{equation}
For the tensors given in Eqs.~(\ref{Eq:AM_Current_Dirac_01}) and (\ref{Eq:AM_Current_EM_01}), and using Maxwell and Dirac equations, one can show that Eq.~(\ref{Eq:Conservation_Law_01}) holds for the total angular momentum tensor. 

For the case that $\mu\nu = ij$, where $i,j=1,2,3$, we find the angular momentum currents due to rotations. We find for the spin and OAM currents of the Dirac field
\begin{subequations}\label{Eq:Div_AM_Dirac}
    \begin{equation}\label{Eq:Div_Spin_Dirac}
        \partial_\lambda \mathcal{S}_\text{D}^{ij,\lambda} = \varepsilon_{ijk} \left[ \hbar \frac{\partial}{\partial t}(\psi^\dagger \bm \Sigma \psi) + \frac{\hbar c}{2}\nabla(\psi^\dagger \gamma^5 \psi)\right]_k,
    \end{equation}
    \begin{equation}\label{Eq:Div_OAM_Dirac}
        \partial_\lambda \mathcal{L}_\text{D}^{ij,\lambda} = \varepsilon_{ijk} \left[ -\hbar \frac{\partial }{\partial t} \mathcal{R}\left\{i\psi^\dagger (\bm r \times \nabla)\psi\right\} - \hbar c \nabla\cdot \mathcal{R}\left\{i\Bar{\psi}\bm \gamma (\bm r\times \nabla)\psi  \right\} \right]_k
    \end{equation}
\end{subequations}
where
\begin{equation}
    \Sigma_i = \frac{1}{2}\varepsilon_{ijk}\sigma^{jk} =  \frac{i}{4} \varepsilon_{ijk} [\gamma^j,\gamma^k] = \frac{1}{2}\begin{pmatrix}
     \sigma_i & 0 \\
    0 &  \sigma_i
    \end{pmatrix},
\end{equation}
with $\sigma_i$ being the Pauli matrices, $\mathcal{R}\{\cdots\}$ takes the real part of its argument, $\gamma^5 = i\gamma^0\gamma^1\gamma^2\gamma^3$ is the chirality operator in the Dirac equation \cite{greiner2013field}, and
\begin{equation}
    \bm \gamma = \gamma^1 \hat{x} + \gamma^2 \hat{y} + \gamma^3 \hat{z}.
\end{equation}
Note that we have used the fact that, using the Dirac equation,
\begin{equation}\label{Eq:Dirac_equation}
    \left( i\hbar\gamma^\mu \partial_\mu - e\gamma^\mu A_\mu - mc\right) \psi = 0,
\end{equation}
we get $\mathfrak{L}_\text{D} = 0$ for the fields that follow Dirac equation of motion. This is straightforward to show by multiplying Eq.~(\ref{Eq:Dirac_equation}) from left by $c\Bar{\psi}$. For the spin and OAM currents of the electromagnetic field we find
\begin{subequations}\label{Eq:Div_AM_EM}
    \begin{equation}\label{Eq:Div_Spin_EM}
        \partial_\lambda \mathcal{S}_\text{em}^{ij,\lambda} = \varepsilon_{ijk}\left[ \epsilon \frac{\partial}{\partial t}(\bm E \times \bm A) - \frac{1}{\mu_0} \nabla\cdot (\bm A\bm B) + \frac{1}{\mu_0} \nabla(\bm A\cdot \bm B)\right]_k
    \end{equation}
    \begin{equation}\label{Eq:Div_OAM_EM}
        \partial_\lambda \mathcal{L}_\text{em}^{ij,\lambda} = \varepsilon_{ijk} \left\{ \epsilon \frac{\partial}{\partial t} \left[\bm E\cdot (\bm r\times \nabla)\bm A\right] - \nabla\cdot \left[\frac{1}{\mu_0} \bm B \times (\bm r \times \nabla)\bm A - \epsilon\bm E(\bm r \times \nabla\phi)\right] + (\bm r \times \nabla)\mathfrak{L}_\text{em}\right\}_k
    \end{equation}
\end{subequations}
The problem with Eqs.~(\ref{Eq:Div_OAM_Dirac}) and (\ref{Eq:Div_AM_EM}) is that they are gauge-dependent, which means that under the transformations $\psi\rightarrow \psi e^{\frac{i}{e}\zeta}$ and $A_\mu \rightarrow A_\mu + \partial_\mu \zeta$ these expressions change. Therefore, the individual terms do not represent any physically meaningful quantity. The fundamental equation to hold is Eq.~(\ref{Eq:Conservation_Law_01}). Therefore, as long as this relation is satisfied, we can cast Eqs.~(\ref{Eq:Div_OAM_Dirac}) and (\ref{Eq:Div_AM_EM}) into gauge-independent forms. To do so, we break $\bm A$ into two longitudinal and transverse parts as $\bm A = \bm A^\parallel + \bm A^\bot$, where $\nabla \cdot \bm A^\bot =0 $ and $\nabla\times \bm A^\parallel = 0$ by definition. After some algebra, we find for the new spin and OAM tensors of Dirac and electromagnetic field
\begin{subequations}\label{Eq:Div_AM_GI}
    \begin{equation}\label{Eq:Div_Spin_Dirac_GI}
        \partial_\lambda \text{S}_\text{D}^{ij,\lambda} = \varepsilon_{ijk} \left[\hbar \frac{\partial}{\partial t}(\psi^\dagger \bm \Sigma \psi) + \frac{\hbar c}{2} \nabla(\psi^\dagger \gamma^5\psi)\right]_k,
    \end{equation}
    \begin{equation}\label{Eq:Div_OAM_Dirac_GI}
        \partial_\lambda \text{L}_\text{D}^{ij,\lambda} = \varepsilon_{ijk} \left[\frac{\partial}{\partial t}\mathcal{R}\left\{ \psi^\dagger (\bm r\times \bm p_\parallel)\psi \right\} + c \nabla\cdot\mathcal{R}\left\{ \Bar{\psi}\bm \gamma (\bm r\times \bm p_\parallel) \psi \right\}\right]_k,
    \end{equation}
    \begin{equation}\label{Eq:Div_Spin_EM_GI}
        \partial_\lambda \text{S}_\text{em}^{ij,\lambda} = \varepsilon_{ijk} \left[\epsilon \frac{\partial}{\partial t}(\bm E\times \bm A^\bot) - \frac{1}{\mu_0}\nabla\cdot (\bm A^\bot \bm B) + \frac{1}{\mu_0}\nabla(\bm A^\bot\cdot\bm B)\right]_k,
    \end{equation}
    \begin{equation}\label{Eq:Div_OAM_EM_GI}
        \partial_\lambda \text{L}_\text{em}^{ij,\lambda} = \varepsilon_{ijk} \left\{\epsilon \frac{\partial }{\partial t }\left[ \bm E\cdot(\bm r \times \nabla)\bm A^\bot \right] - \nabla\cdot\left[ \frac{1}{\mu_0} \bm B \times(\bm r \times \nabla)\bm A^\bot + \epsilon \bm E(\bm r\times \bm E^\parallel) \right] + (\bm r\times\nabla)\mathfrak{L}_\text{em}\right\}_k.
    \end{equation}
\end{subequations}
where $\bm p_\parallel = -i\hbar \nabla - e \bm A^\parallel$ is the gauge-independent covariant momentum operator of electron. We can further separate the contribution of longitudinal electric field $\bm E^\parallel$ to the angular momentum. This contribution can be written as 
\begin{equation}
    \bm E^\parallel\times\bm A^\bot + E_i^\parallel (\bm r \times \nabla) A_i^\bot = - \nabla \times \left[ \bm r (\bm E^\parallel\cdot \bm A^\bot) \right] - \nabla \cdot \left[ \bm A^\bot (\bm r \times \bm E^\parallel) \right]
\end{equation}
When integrated over the entire space, both of these terms on r.h.s of this expression become zero due to the Stokes theorem and thus the longitudinal electric field does not contribute to the global angular momentum of the electromagnetic field. For this reason, we move this term into the angular momentum current terms so that the global angular momentum represents the integrated angular momentum density. Making this change, we get for the new components of the electromagnetic angular momentum currents
\begin{subequations}\label{Eq:Div_AM_GI_E_T}
    \begin{equation}\label{Eq:Div_Spin_EM_GI}
        \partial_\lambda \text{S}_\text{em}^{ij,\lambda} = \varepsilon_{ijk} \left[\epsilon \frac{\partial}{\partial t}(\bm E^\bot \times \bm A^\bot) - \frac{1}{\mu_0}\nabla\cdot (\bm A^\bot \bm B) + \frac{1}{\mu_0}\nabla(\bm A^\bot\cdot\bm B)\right]_k,
    \end{equation}
    \begin{equation}\label{Eq:Div_OAM_EM_GI}
    \begin{split}
        \partial_\lambda \text{L}_\text{em}^{ij,\lambda} = \varepsilon_{ijk} \Bigg\{ & \epsilon \frac{\partial }{\partial t }\left[ \bm E^\bot \cdot(\bm r \times \nabla)\bm A^\bot \right] - \nabla\cdot\left[ \frac{1}{\mu_0} \bm B \times(\bm r \times \nabla)\bm A^\bot + \epsilon_0 \bm A^\bot \left(\bm r\times \frac{\partial \bm E^\parallel}{\partial t}\right)\right] \\
        + & (\bm r\times\nabla)\left[-\frac{1}{2\mu_0} \bm B\cdot \bm B + \frac{\epsilon_0}{2} \bm E^\bot\cdot \bm E^\bot + \epsilon_0 \frac{\partial\bm E^\parallel}{\partial t}\cdot \bm A^\bot \right]\Bigg\}_k.
    \end{split}
    \end{equation}
\end{subequations}
Adding these four equations together we get Eq.~(6) in the main manuscript, 
\begin{equation}\label{Eq:Conservation_Law_02}
    \frac{\partial \bm M}{\partial t} + \nabla_i {J}_{ij} + \nabla\chi + \nabla_i N_{ij} = 0
\end{equation}
where the terms are given in Eqs.~(7) and (8) of the manuscript. Eq.~(\ref{Eq:Conservation_Law_02}) can be written as 
\begin{equation}\label{Eq:Conservation_Law_03}
    \frac{\partial \bm M}{\partial t} + \nabla \cdot \overleftrightarrow{\bm T} = 0
\end{equation}
where
\begin{equation}\label{Eq:J_M}
    \overleftrightarrow{\bm T} = \chi\overleftrightarrow{I} + \overleftrightarrow{\bm J}  + \overleftrightarrow{N}
\end{equation}
with $\overleftrightarrow{I} = \hat{x}\hat{x} + \hat{y}\hat{y} + \hat{z}\hat{z} = \delta_{ij}\hat{x}_i \hat{x}_j$ , and
\begin{equation}
    \overleftrightarrow{N} = \varepsilon_{ijk} \hat{x_i}\hat{x_j} x_k \mathfrak{N}_\text{em}  = \begin{pmatrix}
        0 & z & -y \\
        -z & 0 & x \\
        y & -x & 0
    \end{pmatrix}\mathfrak{N}_\text{em},\quad \mathfrak{N}_{em} = \frac{\epsilon_0}{2} \bm E^\bot\cdot \bm E^\bot -\frac{1}{2\mu_0} \bm B\cdot \bm B + \epsilon_0 \frac{\partial\bm E^\parallel}{\partial t}\cdot \bm A^\bot.
\end{equation}
Note that the first two terms of $\mathfrak{N}_{em}$ describe the Lagrangian due to the transverse electric field, which can be interpreted as the Lagrangian of the free photon \cite{yang2020quantum}, while the last term shows the interaction between the transverse vector potential and the currents due to the longitudinal electric fields. Note that the term $\frac{\epsilon_0}{2}\bm E^\parallel \cdot \bm E^\parallel$ in the Lagrangian of the electromagnetic field cancels out with the same term coming from $\nabla_i J_{ij}$. The expressions for $\overleftrightarrow{\bm J}$ and $\chi$ are given in the main manuscript. 

Also note that the EM Lagrangian does not include the interaction term $j_c^\mu A_\mu$ because it appears in the Dirac part of the Lagrangian $\mathfrak{L}_\text{D}$. Dirac Lagrangian, as mentioned earlier, vanishes from the expression for the OAM of the Dirac field since it satisfies the Dirac equation. For this reason, the contribution from the Dirac Lagrangian $\mathfrak{L}_{\text{D}}$ disappears from the conservation equations. 

Equation~(\ref{Eq:Conservation_Law_03}) describes the local conservation law for the angular momentum currents $\overleftrightarrow{\bm T}$ and the angular momentum density (charge) $\bm M$. When integrated over the entire space, and assuming that the fields vanish on the boundary of the this surface, the second term becomes an integral over this surface and thus vanishes. In this case, we arrive at the usual global conservation of angular momentum equation which states that the total angular momentum of the Dirac and Maxwell fields is a constant. However, in situations where the problem under consideration is an open dissipative system, this simplification cannot be made and surface terms of the angular momentum current can carry angular momentum out of the system. 

\section{Spin-Orbit Torque}
In this section we show that Eq.~(\ref{Eq:Div_AM_GI}) leads to Eq.~(9) in the main manuscript. Starting with the equation for the spin of the Dirac fields, we get for the $z$ component for instance, 
\begin{equation}
    \partial_\lambda \text{S}_\text{D}^{12,\lambda} = \frac{\hbar}{2} \partial_t (\psi^\dagger \Sigma_z \psi) + \frac{\hbar c}{2} \partial_z (\psi^\dagger \gamma^5 \psi)
\end{equation}
where $\partial_t \equiv \frac{\partial}{\partial t}$ and $\partial_i \equiv \frac{\partial}{\partial x_i}$. Using Dirac equation we get 
\begin{subequations}\label{Eq:Dirac_Equation}
  \begin{equation}
      \hbar \partial_t \psi = -\hbar c \gamma^0 (\bm \gamma \cdot \bm\nabla\psi) - ice A_\mu \gamma^0 \gamma^\mu \psi - im c^2 \gamma^0 \psi 
  \end{equation}
  \begin{equation}
    \hbar \partial_t \psi^\dagger = \hbar c(\bm \nabla \psi^\dagger \cdot \bm \gamma)\gamma^0 + ice A_\mu \psi^\dagger \gamma^0 \gamma^\mu + im c^2 \psi^\dagger \gamma^0
  \end{equation}
\end{subequations}
Using these equations we find
\begin{equation}\label{Eq:SO_Derivation_Spin_Dirac}
    \begin{split}
        \partial_\lambda \text{S}_\text{D}^{12,\lambda} = & \frac{\hbar}{2}\left\{ (\partial_t \psi^\dagger) \Sigma_z \psi + \psi^\dagger \Sigma_z (\partial_t \psi) + c \partial_z (\psi^\dagger\gamma^5 \psi) \right\}\\
        + & \frac{1}{2} \Big\{ \hbar c (\bm \nabla\psi^\dagger \cdot \bm \gamma ) i\gamma^0\gamma^1\gamma^2 \psi - ce A_\mu \psi^\dagger \gamma^0 \gamma^\mu \gamma^1 \gamma^2 \psi - mc^2 \psi^\dagger \gamma^0 \gamma^1\gamma^2 \psi \\
        - & i \hbar c \psi^\dagger \gamma^1 \gamma^2 \gamma^0 (\bm \gamma\cdot \bm \nabla \psi) + ce A_\mu \psi^\dagger \gamma^1 \gamma^2 \gamma^0 \gamma^\mu \psi + mc^2 \psi^\dagger \gamma^1 \gamma^2 \gamma^0 \psi + \hbar c \partial_z (\psi^\dagger \gamma^5 \psi) \Big\} \\
        = & \frac{1}{2}\Big\{\! \!-\! \hbar c \partial_z (\psi^\dagger \gamma^5 \psi)\! + \! i \hbar c (\partial_x \psi^\dagger) \gamma^0 \gamma^2 \psi - i \hbar c \psi^\dagger \gamma^0 \gamma^2 (\partial_x \psi) - i \hbar c (\partial_y \psi^\dagger) \gamma^0 \gamma^1 \psi + i \hbar c \psi^\dagger \gamma^0 \gamma^1 (\partial_y \psi ) \\
        + & 2 ce A_1 \psi^\dagger \gamma^0 \gamma^2 \psi - 2 ce A_2 \psi^\dagger \gamma^0 \gamma^1 \psi + \hbar c \partial_z (\psi^\dagger \gamma^5 \psi )\Big\} \\
        = & \hbar c \mathcal{R}\left\{ i \bar{\psi} (\gamma^1 \partial_y - \gamma^2 \partial_x) \psi \right\} - ce(A_x^\parallel \bar{\psi}\gamma^2\psi - A_y^\parallel \bar{\psi}\gamma^1 \psi)  - (A_x^\bot j_{c,y} - A_y^\bot j_{c,x}) \\
        = & - c \mathcal{R}\left\{ \bar{\psi}\left[ \gamma^1(-i\hbar\partial_y - e A_y^\parallel) - \gamma^2(-i\hbar\partial_x - e A_x^\parallel) \right]\psi \right\}- (A_x^\bot j_{c,y} - A_y^\bot j_{c,x}) \\
        = &  - c \mathcal{R}\left\{ \bar{\psi}(\bm \gamma \times \bm p_\parallel)_z \psi \right\} + (\bm j_c \times \bm A^\bot )_z.
    \end{split}
\end{equation}
In this derivation we have used the facts that $\left\{\gamma^\mu,\gamma^\nu\right\} = 2\eta^{\mu\nu}$, $(\gamma^i)^2 = -1$, $A_1 = - A_x$, $A_2 = - A_y$, and $\bm j_c = ce \bar{\psi}\bm \gamma \psi$. We can do a similar derivation for other components of $\partial_\lambda \text{S}_\text{D}^{ij,\lambda}$. Doing so we find Eq.~(9a) in the main manuscript. Note that the first term on the r.h.s. of Eq.~(\ref{Eq:SO_Derivation_Spin_Dirac}) is nothing but the spin-orbit torque of the Dirac field $\bm \tau_\text{D}$ given in Eq.~(10) of the paper. 

We now turn into the equation for the OAM of the Dirac field Eq.~(\ref{Eq:Div_OAM_Dirac_GI}). We get for the $z$ component, for instance

\begin{equation}
    \begin{split}
        \partial_\lambda \text{L}_\text{D}^{12,\lambda} = & \partial_t \mathcal{R}\left\{ \psi^\dagger (x p_{\parallel,y} - y p_{\parallel,x})\psi \right\} + c \bm\nabla \cdot \mathcal{R}\left\{ \bar{\psi} \bm \gamma (x p_{\parallel,y} - y p_{\parallel,x})\psi \right\} \\
        = & \mathcal{R}\Big\{ (\partial_t \psi^\dagger) (x p_{\parallel,y} - y p_{\parallel,x}) \psi + \psi^\dagger (x p_{\parallel,y} - y p_{\parallel,x}) (\partial_t \psi) + c \bm\nabla \cdot \bar{\psi} \bm \gamma (x p_{\parallel,y} - y p_{\parallel,x})\psi \\
        - & e \psi^\dagger \psi (x\partial_t A_y^\parallel - y\partial_t A_x^\parallel) \Big\} \\
        = & \mathcal{R}\Big\{ c (\bm \nabla \psi^\dagger \cdot \bm \gamma)\gamma^0  (x p_{\parallel,y} \!-\! y p_{\parallel,x})  \psi +\! \frac{i c e}{\hbar} A_\mu \psi^\dagger \gamma^0 \gamma^\mu  (x p_{\parallel,y}\!\! - \! y p_{\parallel,x}) \psi + \!\frac{imc^2}{\hbar} \psi^\dagger \gamma^0  (x p_{\parallel,y}\! - \! y p_{\parallel,x}) \psi \\
        - & c \psi^\dagger (x p_{\parallel,y}\! - \! y p_{\parallel,x}) \gamma^0 (\bm \gamma\cdot \bm \nabla \psi) - \frac{ice}{\hbar} \psi^\dagger (x p_{\parallel,y}\! - \! y p_{\parallel,x}) \gamma^0 \gamma^\mu (A_\mu \psi) - \frac{imc^2}{\hbar} \psi^\dagger (x p_{\parallel,y}\! - \! y p_{\parallel,x}) \gamma^0 \psi \\
        + & c \bm \nabla \cdot \bar{\psi} \bm \gamma (x p_{\parallel,y} - y p_{\parallel,x})\psi - e \psi^\dagger \psi (x\partial_t A_y^\parallel - y\partial_t A_x^\parallel) \Big\} \\
        = & \mathcal{R}\Big\{\!\!-\! c \bm \nabla \cdot \bar{\psi} \bm \gamma (x p_{\parallel,y} - y p_{\parallel,x}) \psi + c \bar{\psi} \left[\bm \gamma \cdot \bm \nabla (x p_{\parallel,y}\! - \! y p_{\parallel,x})\right] \psi - ce \bar{\psi}\gamma^\mu \psi (x\partial_y - y \partial_x)A_\mu \\
        + & c \bm \nabla \cdot \bar{\psi} \bm \gamma (x p_{\parallel,y} - y p_{\parallel,x})\psi - e \psi^\dagger \psi (x\partial_t A_y^\parallel - y\partial_t A_x^\parallel) \Big\} \\
        = & \mathcal{R}\left\{ c \bar{\psi} (\gamma^1 p_{\parallel,y} - \gamma^2 p_{\parallel,x})\psi \right\} - ce \bar{\psi}\bm \gamma \psi (x \bm\nabla A_y^\parallel - y\bm\nabla A_x^\parallel) - ce \bar{\psi}\gamma^\mu \psi (x\partial_y - y \partial_x)A_\mu \\
        - & e \psi^\dagger \psi (x\partial_t A_y^\parallel - y\partial_t A_x^\parallel) \\
        = & \mathcal{R}\left\{ c \bar{\psi} (\gamma^1 p_{\parallel,y} - \gamma^2 p_{\parallel,x})\psi \right\} - ce \bar{\psi} \Big[ \gamma^1 (x\partial_x A_y^\parallel - y \partial_x A_x^\parallel) + \gamma^2 (x \partial_y A_y^\parallel - y \partial_y A_x^\parallel) \\
        + & \gamma^3 (x \partial_z A_y^\parallel - y \partial_z A_x^\parallel) - \gamma^1 (x\partial_y A_x - y\partial_x A_x) - \gamma^2 (x\partial_y A_y - y\partial_x A_y) - \gamma^3 (x\partial_y A_z - y\partial_x A_z) \\
        + & \frac{1}{c}\gamma^0 (x\partial_y \phi - y\partial_x \phi) + \frac{1}{c} \psi^\dagger (x\partial_t A_y^\parallel - y\partial_t A_x^\parallel) \Big]\psi \\
        = & \mathcal{R}\left\{ c \bar{\psi} (\gamma^1 p_{\parallel,y} - \gamma^2 p_{\parallel,x})\psi \right\} - ce \bar{\psi}\gamma^1 \psi x (\partial_x A_y^\parallel - \partial_y A_x^\parallel) + ce\bar{\psi}\gamma^1 \psi (x\partial_y - y\partial_x)A_x^\bot \\
        - & ce \bar{\psi}\gamma^2 \psi y (\partial_x A_y^\parallel - \partial_y A_x^\parallel) + ce\bar{\psi}\gamma^2 \psi (x\partial_y - y\partial_x)A_y^\bot - ce \bar{\psi}\gamma^3 \psi x(\partial_z A_y^\parallel - \partial_y A_z^\parallel) \\ 
        + & ce \bar{\psi}\gamma^3\psi y (\partial_z A_x^\parallel \! -\! \partial_x A_z^\parallel) +\! ce \bar{\psi}\gamma^3\psi (x\partial_y \!-\!\!\! y\partial_x)A_z^\bot + e \psi^\dagger \psi \left[x (-\partial_y\phi\! -\! \partial_t A_y^\parallel) \!- \!y (-\partial_x \phi - \partial_t A_x^\parallel)\right] \\
        = & ~ c \mathcal{R}\left\{ \bar{\psi} (\bm \gamma \times \bm p_\parallel)_z \psi  \right\}  + j_{c,i}(\bm r\times \bm \nabla)_z A_i^\bot + \rho (\bm r \times \bm E^\parallel)_z
        \end{split}
\end{equation}
where we have used the facts that $\bm\nabla\times \bm A^\parallel = 0 $ and $\bm E^\parallel = -\bm\nabla \phi - \partial_t \bm A^\parallel$. Writing similar equations for the other components we get Eq.~(9b) in the main manuscript. 

We can repeat this derivation for the electromagnetic spin and OAM currents as well. Using Eq.~(\ref{Eq:Div_Spin_EM_GI}) we find, again for the $z$ component for instance,
\begin{equation}
    \begin{split}
        \partial_\lambda \text{S}_\text{em}^{12,\lambda } = & \epsilon_0 \partial_t (\bm E^\bot \times \bm A^\bot)_z - \frac{1}{\mu_0} \bm\nabla\cdot(\bm A^\bot B_z) + \frac{1}{\mu_0}\partial_z (\bm A^\bot \cdot \bm B) \\ 
        = & \epsilon_0 \partial_t (\bm E^\bot \times\bm A^\bot)_z -\frac{1}{\mu_0} B_z (\bm\nabla\cdot\bm A^\bot) - \frac{1}{\mu_0} (\bm A^\bot \cdot \bm\nabla)B_z \\
        + &  \frac{1}{\mu_0} \left\{ (\bm A^\bot \cdot \bm\nabla)B_z + (\bm B\cdot \bm\nabla)A_z^\bot + [\bm A^\bot \times (\bm\nabla\times \bm B)]_z + [\bm B\times (\bm\nabla\times \bm A^\bot)]_z  \right\} \\
        = & \epsilon_0 \partial_t(\bm E^\bot \times \bm A^\bot) + \frac{1}{\mu_0} (\bm B \cdot \bm\nabla)A_z^\bot - \epsilon_0 [ (\partial_t \bm E)\times \bm A^\bot ]_z - (\bm j_c \times \bm A^\bot)_z \\
        = & \epsilon_0 [\bm E^\bot \times (\partial_t \bm A^\bot)]_z + \frac{1}{\mu_0}(\bm B \cdot \bm\nabla)A_z^\bot - \epsilon_0 \left[ \frac{\partial \bm E^\parallel}{\partial t}\times \bm A^\bot \right] - (\bm j_c\times \bm A^\bot)_z \\
        = & \frac{1}{\mu_0}(\bm B \cdot \bm\nabla)A_z^\bot - \epsilon_0 \left(\frac{\partial \bm E^\parallel}{\partial t} \times \bm A^\bot \right)_z  - (\bm j_c\times \bm A^\bot)_z ,
    \end{split}
\end{equation}
where we have used the facts that $\bm\nabla\cdot \bm A^\bot =0$, $\bm B = \bm\nabla\times \bm A^\bot$, $\bm E^\bot = -\partial_t \bm A^\bot$, and $\bm \nabla\times \bm B = \frac{1}{c^2} \partial_t \bm E + \mu_0 \bm j_c$ . By repeating this derivation for the other components we get Eq.~(9c) in the paper. Note that the first two terms are the spin-orbit torque of the electromagnetic field, $\bm\tau_\text{em}$, given in Eq.~(11) of the main manuscript. 

We can repeat this derivation to find the equation for $\partial_\lambda \text{L}^{ij,\lambda}_\text{em}$. However, using the continuity condition,
\begin{equation}
    \partial_\lambda (\text{S}_\text{D}^{ij,\lambda} + \text{L}_\text{D}^{ij,\lambda} + \text{S}_\text{em}^{ij,\lambda} + \text{L}_\text{em}^{ij,\lambda}) = 0 ,
\end{equation}
it is straightforward to show that Eq.~(9d) of the main manuscript holds. 

\section{\textit{Semi-Local} Conservation Law}
We can get the semi-local conservation laws by integrating the terms in Eqs.~(\ref{Eq:Div_Spin_Dirac_GI}), (\ref{Eq:Div_OAM_Dirac_GI}), and (\ref{Eq:Div_AM_GI_E_T}) over the volume $V'$, on which surface the Dirac eigenfunctions $\psi$ become zero. Using Gauss's theorem, the integral of the terms $\nabla(\psi^\dagger \gamma^5 \psi)$ and $\nabla \cdot \mathcal{R}\{ \Bar{\psi} \bm \gamma (\bm r \times \bm p)\psi \}$ vanish because they become surface integrals of functions of $\psi$. We therefore arrive at the semi-local conservation law
\begin{equation}\label{Eq:Semi-Local_Conservation}
    \frac{\partial \tilde{\bm M}_\text{D}}{\partial t} + \frac{\partial \tilde{\bm M}_\text{em}}{\partial t} + \tilde{{\bm{J}}}_\text{A} + \tilde{\bm h} + \int_{V'} \nabla\times(\bm r \mathfrak{N}_\text{em})dV' = 0
\end{equation}
where $\tilde{\bm M}_\text{D}$ is the total angular momentum of the electron given by
\begin{equation}\label{Eq:Conservation_law_Semi-local}
    \tilde{\bm M}_\text{D} = \hbar \int_{V'}\left[\frac{1}{2}(\psi^\dagger \bm \Sigma \psi) + \mathcal{R} \{ \psi^\dagger (\bm r \times \bm p)\psi \} \right]dV',
\end{equation}
$\tilde{\bm M}_\text{em}$ is the EM total angular momentum in the volume $V'$,
    \begin{equation}\label{Eq:TAM_EM_Semi}
        \Tilde{\bm M}_\text{em} = \int_{V'} \epsilon(\bm E^\bot \times \bm A^\bot)d^3x
        + \int_{V'} \epsilon\left[ \bm E^\bot \cdot (\bm r\times \nabla)\bm A^\bot \right]d^3x,
    \end{equation}
and
\begin{subequations}
  \begin{equation}\label{Eq:Semi-local_AM_Current}
    \tilde{{\bm J}}_\text{A} = - \int_{V'} \nabla \cdot \left[\frac{1}{\mu_0}\bm A^\bot \bm B + \frac{1}{\mu_0} \bm B \times (\bm r \times \nabla)\bm A^\bot + \epsilon \bm A^\bot \left(\bm r \times \frac{\bm \partial E^\parallel}{\partial t}\right) \right]dV',
\end{equation}
\begin{equation}\label{Eq:Semi-Local_Helicity}
    \tilde{\bm h} = \int_{V'} \nabla(\bm A^\bot \cdot \bm B) dV'.
\end{equation}
\end{subequations}
Using Gauss's theorem the volume integrals in Eqs.~(\ref{Eq:Semi-Local_Conservation}), (\ref{Eq:Semi-local_AM_Current}) and (\ref{Eq:Semi-Local_Helicity}) can be converted into surface integrals. We find
\begin{subequations}\label{Eq:Terms_Semi-local}
\begin{equation}\label{Eq:TAM_Currents_Semi-local}
        \Tilde{\bm J}_\text{A} = - \int_{S'} \hat{n}_i \Big[ A_i^\bot \bm B + \epsilon_{ijk} B_j (\bm r \times \nabla) A_k^\bot
        + \epsilon A^\bot_i \left(\bm r \times \frac{\bm \partial E^\parallel}{\partial t}\right)  \Big] da
\end{equation}

    \begin{equation}\label{Eq:helicity_Semi-local}
        \Tilde{\bm h} = \int_{S'} \hat{\bm n} (\bm A^\bot \cdot \bm B) da
    \end{equation}
\end{subequations}
where $\hat{\bm n}$ is the unit vector normal to the surface of the volume $V'$, and $da$ its surface element. Equation~(\ref{Eq:Semi-Local_Conservation}) presents an equation for the time evolution of angular momentum of the electron in terms of the EM fields.

\section{Symmetrized Angular Momentum Tensor}
In our derivation, we have used the canonical form of the angular momentum tensor which is derived directly from the application of Noether theorem to the QED Lagrangian in Eq.~(\ref{Eq:Dirac_lagrangian}). The angular momentum tensor can be written in terms of the canonical energy-momentum tensor, $T^{\mu\nu}$, as
\begin{equation}
    \mathcal{M}^{\mu\nu,\lambda} = x^\mu T^{\lambda\nu} - x^\nu T^{\lambda\mu} + S^{\mu\nu,\lambda}.
\end{equation}
Conservation of angular momentum implies that, when $\partial_\lambda S^{\mu\nu,\lambda}\neq 0$, 
\begin{equation}
    T^{\mu\nu} \neq T^{\nu\mu}
\end{equation}
This in unpleasant because in general relativity, the energy-momentum tensor is directly proportional to the metric tensor which is symmetric in $\mu$ and $\nu$. To overcome this problem, the energy momentum tensor can be modified to the so-called Bellifante-Resenfeld energy-momentum tensor as \cite{belinfante1940current,rosenfeld1940tenseur} 
\begin{equation}
    {T}^{'\mu\nu} = T^{\mu\nu} + \frac{1}{2} \partial_\lambda(S^{\nu\lambda,\mu} + S^{\mu\lambda,\nu} - S^{\nu\mu,\lambda})
\end{equation}
It can be shown that this new energy-momentum tensor is symmetric and does not change the conservation law of the energy-momentum tensor, $\partial_\mu {T}^{'\mu\nu} = 0$. We can therefore write a new symmetrized angular momentum tensor, ${M}^{'\mu\nu,\lambda}$ as 
\begin{equation}
    \mathcal{M}^{'\mu\nu,\lambda} = x^\mu {T}^{'\lambda\nu} - x^\nu {T}^{'\lambda\mu}
\end{equation}
where we do not need to include the additional spin tensor because it is already present in the symmetric energy-momentum tensor. This symmetrized angular momentum tensor is of course different from the canonical one we derived in Eqs.~(\ref{Eq:AM_Current_Dirac_01}) and (\ref{Eq:AM_Current_EM_01}). However, it is still not gauge invariant and the conservation law of angular momentum still holds. In other words
\begin{equation}
    \partial_\lambda \mathcal{M}^{'\mu\nu,\lambda} = \partial_\lambda \mathcal{M}^{\mu\nu,\lambda} = 0
\end{equation}
We can follow a similar procedure as we did in the previous section to derive the expression for the gauge-independent forms of the symmetrized angular momentum tensor for the electromagnetic and Dirac fields. We find, setting $\mu,\nu = i,j$,
\begin{equation}\label{Eq:Conservation_Law_Symm}
    \frac{\partial \bm M^{'}}{\partial t} + \nabla\cdot \overleftrightarrow{\bm J^{'}} + \nabla{\chi'} - \nabla\times(\bm r \mathcal{U}_\text{em}) = 0
\end{equation}
where
\begin{subequations}
    \begin{equation}
        {\bm M^{'}} = \frac{\hbar}{2} \left(\psi^\dagger \bm \Sigma \psi \right) + \mathcal{R}\left\{ \psi^\dagger (\bm r\times \bm p_\parallel)\psi \right\} + \epsilon_0~ \bm r \times(\bm E\times\bm B) - \rho(\bm r\times \bm A^\bot)
    \end{equation}
    \begin{equation}
        \nabla \cdot {\overleftrightarrow{\bm J^{'}}} = c \nabla_i \mathcal{R}\{ \Bar{\psi} \gamma^i (\bm r\times \bm p_\parallel)\psi \} + \epsilon_0~ \nabla_i  E_i (\bm r\times \bm E) +  \frac{1}{\mu_0} \nabla_i \left[ B_i(\bm r\times \bm B)\right]- \nabla_i \left[ J_i(\bm r \times \bm A^\bot)\right]
    \end{equation}
    \begin{equation}
        \tilde{\chi} = \frac{\hbar c}{2}(\psi^\dagger \gamma^5 \psi)
    \end{equation}
    \begin{equation}
        \mathcal{U}_\text{em} = \frac{1}{2} \left( \epsilon_0 \bm E\cdot \bm E + \frac{1}{\mu_0} \bm B\cdot \bm B \right)
    \end{equation}
\end{subequations}
We emphasize that Eq.~(\ref{Eq:Conservation_Law_Symm}) is identical to the conservation Eq.~(6) in the main text. In fact, the terms related to the Dirac field are exactly the same as the one with the four-divergence of the canonical angular momentum tensor. The main difference is that we lose separate physically observable expressions for the spin and OAM densities and currents of the electromagnetic field and instead we get expressions for the total angular momentum,
\begin{equation}
    \epsilon_0 \bm r\times (\bm E\times\bm B) - \rho (\bm r\times \bm A^\bot),
\end{equation}
and total angular momentum currents,
\begin{equation}
    \epsilon_0~ \bm E (\bm r\times \bm E) + \frac{1}{\mu_0} \bm B (\bm r\times \bm B) - \bm J(\bm r\times \bm A^\bot),
\end{equation}
of the electromagnetic field. Assuming that the symmetric energy-momentum tensor is more fundamental than the canonical one, however, does not prevent us from writing conservation laws that involve canonical angular momentum tensor and taking its gauge-independent terms as physically meaningful quantities.

\section{Boost Conservation Relations}
We started our derivation by applying the Noether's theorem to the Dirac-Maxwell fields under the Lorentz transformation. To derive the angular momentum conservation equations, however, we only focused on the spatial rotations of the coordinates i.e. the space-space components of the angular momentum tensor current $\mathcal{M}^{\mu\nu,\lambda}$. In this section, we look at the conservation equations for the time-space components of the Lorentz transformations, namely the boosts, of the Dirac-Maxwell fields. 

Using a similar approach to the one used in the first section, the conservation equation resulting from the boost components of the Lorentz transformation is given by, 

\begin{equation}\label{Eq:Div_AM0_Total_GI}
    \begin{split}
       \partial_\lambda \mathcal{M}^{i0,\lambda}  = & \partial_\lambda (\mathcal{M}_\text{D})^{i0,\lambda} + \partial_\lambda (\mathcal{M}_\text{E})^{i0,\lambda} \\
        = & \frac{\partial}{\partial t}\mathcal{R}\left\{i\psi^\dagger\left[\frac{\bm r}{c}p_0 - ct \bm p_\parallel \right]\psi\right\}  -c\nabla\times (\frac{\hbar}{2}\psi^\dagger\bm \Sigma\psi) + c\nabla.\mathcal{R}\left\{ i\bar{\psi}\bm \gamma\left[\frac{\bm r}{c}p_0 - ct \bm p_\parallel \right]\psi \right\} \\
        - & \epsilon\frac{\partial}{\partial t} \left[\bm E. (\frac{\bm r}{c}\partial_t + ct \nabla)\bm A^\bot\right] +  \frac{1}{\mu_0}\nabla.\left[\bm B\times(\frac{\bm r}{c}\partial_t + ct \nabla)\bm A^\bot\right] + \frac{1}{\mu_0 c} \bm B\times \bm E^\bot + \frac{1}{\mu_0 c} \bm E \cdot (\nabla )\bm A^\bot \\
        + & c\rho \bm A^\bot   -\left(\frac{\bm r}{c}\right)\bm J.\bm E^\parallel + ct \rho \bm E^\parallel + \epsilon \bm E.\left(\frac{\bm r}{c}\partial_t + ct\nabla\right)\bm E^\parallel + c\left(\frac{\bm r}{c^2}\partial_t + t\nabla\right)\mathcal{L}_\text{E} = 0
    \end{split}
\end{equation}

where $p_0 = i\hbar \partial_t - e\phi$ and $\bm p_\parallel = -i\hbar \nabla - e\bm A^\parallel$ are the gauge-independent time-derivative and momentum operators of the Dirac field, respectively. It is a matter of straightforward algebra to show that $\bm r \times \partial_\lambda \bm M^{0,\lambda} = \epsilon_{ijk} x_j \partial_\lambda \mathcal{M}^{k0,\lambda}$ gives the conservation equation of the angular momentum in Eq.~(\ref{Eq:Conservation_Law_03}). 

\section{Two Plane Wave Interference}
We now evaluate the terms in Eq.~(\ref{Eq:Div_Spin_EM}) for the interference of two plane waves at different frequencies. For a plane wave propagating along $\bm k/|\bm k|$, with the wavevector $k$ we have $\bm k \cdot \bm E = \bm k \cdot \bm A^\bot = \bm k \cdot \bm B = 0$. Therefore we get
\begin{equation}
    \nabla \cdot (\bm A^\bot \bm B) = (\bm k \cdot \bm A^\bot)\bm B = 0.
\end{equation}
Note also that the EM spin-orbit torque $\bm \tau_\text{em}$ also vanishes for planes waves because
\begin{equation}
    (\bm B \cdot \nabla)\bm A^\bot = (\bm B \cdot\bm k)\bm A^\bot = 0.
\end{equation}
Thus the only relevant terms in finding the conservation law for the spin current of the two plane wave interference are the time-derivative of spin and gradient of helicity. 

The electric field for two plane waves propagating along $z$ direction can be written as
\begin{equation}
    \bm E = \mathcal{R}\{ \mathbfcal{\bm E}_1 e^{-i(\omega_1 t - k_1 z)} + \mathbfcal{\bm E}_2 e^{-i(\omega_2 t - k_2 z)}  \}
\end{equation}
where $\mathbfcal{E}_i = \frac{\mathcal{E}_i}{\sqrt{2}}(\hat{x} + i \hat{y})$ are the complex electric field amplitudes of the two modes with frequencies $\omega_i/c = k_i $. Using Maxwell equation $\nabla\times \bm E  = - \frac{\partial \bm B}{\partial t}$ and $\bm E = -\frac{\partial \bm A}{\partial t}$, we get
\begin{subequations}
  \begin{equation}
    \bm B = \mathcal{R}\left\{ -\frac{k_1}{i\omega_1}\mathbfcal{\bm E}_1 e^{-i(\omega_1 t - k_1 z)} - \frac{k_2}{i \omega_2} \mathbfcal{\bm E}_2 e^{-i(\omega_2 t - k_2 z)} \right\}
  \end{equation}
  \begin{equation}
    \bm A^\bot = \mathcal{R}\left\{\frac{1}{i\omega_1}\mathbfcal{\bm E}_1 e^{-i(\omega_1t - k_1 z)} + \frac{1}{i\omega_2}\mathbfcal{\bm E}_2e^{-i(\omega_2t - k_2 z)}\right\}
  \end{equation}
\end{subequations}
Using these equations we get for the spin of the two plane waves
\begin{equation}
  \epsilon (\bm E^\bot \times \bm A^\bot) = \frac{\epsilon}{2}\left[\frac{1}{\omega_1} \mathcal{I}\{ \mathbfcal{E}_1^*\times \mathbfcal{E}_1 \} + \frac{1}{\omega_2} \mathcal{I}\{ \mathbfcal{E}^*_2\times\mathbfcal{E}_2\} + \left(\frac{1}{\omega_1}+\frac{1}{\omega_2}\right)\mathcal{I}\left\{ \mathbfcal{E}^*_1 \times \mathbfcal{E}_2 e^{+i[(\omega_1 -\omega_2)t - (k_1 - k_2)z]}\right\} \right]
\end{equation}
and thus we get
\begin{equation}
\begin{split}
  \epsilon\frac{\partial}{\partial t} (\bm E^\bot\times \bm A^\bot) =  &-  \epsilon\frac{\omega_1^2 - \omega_2 ^2}{2\omega_1\omega_2}\mathcal{R}\left\{ \mathbfcal{E}_1^*\times \mathbfcal{E}_2 e^{+i[(\omega_1 - \omega_2)t - (k_1 - k_2)z]}\right\} \\
  = &  -\epsilon \frac{\omega_1^2- \omega_2^2}{2\omega_1\omega_2} \mathcal{I}\left\{ \mathcal{E}_1 \mathcal{E}_2^* e^{-i[(\omega_1-\omega_2)t - (k_1 - k_2)z]} \right\} \hat{z}
\end{split}
\end{equation}
For the helicity density we find
\begin{equation}
  \bm A^\bot \cdot \bm B = -\frac{k_1}{\omega_1^2}|\mathcal{E}_1|^2 - \frac{k_2}{\omega_2^2}|\mathcal{E}_2|^2 - \frac{k_1 + k_2}{2\omega_1 \omega_2}\mathcal{R}\left\{ \mathcal{E}_1 \mathcal{E}_2^* e^{-i[(\omega_1 - \omega_2)t - (k_1 - k_2)z]} \right\}
\end{equation}
and thus
\begin{equation}
\begin{split}
  \frac{1}{\mu_0}\nabla(\bm A^\bot \cdot \bm B) = & \frac{1}{\mu_0}\frac{k_1^2 - k_2^2}{2\omega_1\omega_2}\mathcal{I}\left\{ \mathcal{E}_1\mathcal{E}_2^* e^{-i[(\omega_1 - \omega_2)t - (k_1 - k_2)z]} \right\}\hat{z} \\
  = & \epsilon \frac{\omega_1^2 - \omega_2 ^2}{2\omega_1 \omega_2} \mathcal{I}\left\{ \mathcal{E}_1\mathcal{E}_2^* e^{-i[(\omega_1 - \omega_2)t - (k_1 - k_2)z]} \right\}
\end{split}
\end{equation}
which confirms the conservation Eq.~(14) in the main manuscript. 

\section{Meaning of the Notations}
Throughout this paper, the usual expressions for dot and cross product are assumed. Terms like $\bm A^\bot \bm B$ are tensorial expressions which can be expanded as
\begin{equation}
\begin{split}
    \bm A^\bot \bm B = & A_x^\bot B_x \hat{x}\hat{x} + A_x^\bot B_y \hat{x}\hat{y} + A_x^\bot B_z \hat{x}\hat{z} \\
    + & A_y^\bot B_x \hat{y}\hat{x} + A_y^\bot B_y \hat{y}\hat{y} + A_y^\bot B_z \hat{y}\hat{z} \\
    + & A_z^\bot B_x \hat{z}\hat{x} + A_z^\bot B_y \hat{z}\hat{y} + A_z^\bot B_z \hat{z}\hat{z}
\end{split}
\end{equation}
Similar expressions can be written the terms like $\bm E(\bm r\times \bm E)$, $\bm E(\bm r\times \bm E)$, $\bm J(\bm r\times \bm A^\bot)$, and so on. Therefore, the expression $\nabla\cdot (\bm A^\bot \bm B)$ means
\begin{equation}\label{Eq:Div_of_Tensor}
    \begin{split}
        \nabla\cdot(\bm A^\bot \bm B) = & \left[\partial_x(A_x^\bot B_x) + \partial_y(A_y^\bot B_x) + \partial_z(A_z^\bot B_x)\right]\hat{x} \\
        + & \left[\partial_x(A_x^\bot B_y) + \partial_y(A_y^\bot B_y) + \partial_z(A_z^\bot B_y)\right]\hat{y} \\
        + & \left[\partial_x(A_x^\bot B_z) + \partial_y(A_y^\bot B_z) + \partial_z(A_z^\bot B_z)\right]\hat{x}
    \end{split}
\end{equation}
which is a vector. We can similar expand
\begin{equation}
    \begin{split}
        \bm B\times(\bm r\times\nabla)\bm A^\bot = & \left[B_y (y\partial_z - z\partial_y) A^\bot_z - B_z (y\partial_z - z\partial_y) A^\bot_y \right]\hat{x}\hat{x} \\
        + & \left[B_z (y\partial_z - z\partial_y) A^\bot_x - B_x (y\partial_z - z\partial_y) A^\bot_z \right]\hat{y}\hat{x} \\
        + & \left[B_x (y\partial_z - z\partial_y) A^\bot_y - B_y (y\partial_z - z\partial_y) A^\bot_x \right]\hat{z}\hat{x}  \\
        + & \left[B_y (z\partial_x - x\partial_z) A^\bot_z - B_z (z\partial_x - x\partial_z) A^\bot_y \right]\hat{x}\hat{y} \\
        + & \left[B_z (z\partial_x - x\partial_z) A^\bot_x - B_x (z\partial_x - x\partial_z) A^\bot_z \right]\hat{y}\hat{y} \\
        + & \left[B_x (z\partial_x - x\partial_z) A^\bot_y - B_y (z\partial_x - x\partial_z) A^\bot_x \right]\hat{z}\hat{y} \\
        + & \left[B_y (x\partial_y - y\partial_x) A^\bot_z - B_z (x\partial_y - y\partial_x) A^\bot_y \right]\hat{x}\hat{z} \\
        + & \left[B_z (x\partial_y - y\partial_x) A^\bot_x - B_x (x\partial_y - y\partial_x) A^\bot_z \right]\hat{y}\hat{z} \\
        + & \left[B_x (x\partial_y - y\partial_x) A^\bot_y - B_y (x\partial_y - y\partial_x) A^\bot_x \right]\hat{z}\hat{z}
    \end{split}
\end{equation}
and we can take its divergence similar to Eq.~(\ref{Eq:Div_of_Tensor}).

\vspace{2cm}
\Urlmuskip=0mu plus 1mu\relax
\bibliographystyle{iopart-num}
\bibliography{main}

\end{document}